\documentclass[journal,nopub]{preprint}

\usepackage{fancybox}
\usepackage{curves}
\usepackage{caption}
\usepackage{epstopdf}
\usepackage{epsfig}
\usepackage{psfrag}
\usepackage{amsmath,bm}
\usepackage{amssymb}
\usepackage{graphicx}
\usepackage{array}
\usepackage{booktabs}
\usepackage{tabularx}
\usepackage{subfigure}
\usepackage{colortbl}
\usepackage{xcolor}
\usepackage{multirow}
\colorlet{lblue}{blue!50!black}

\title{A Residual Learning Approach for Unsteady Aerodynamic Load Prediction}

\author[1,*]{Divya Sanghi}
\author[1,*]{Carlos E. S. Cesnik}
\affil[1]{University of Michigan, Ann Arbor, MI, 48109}
\papercitation{
    Sanghi, D. and Cesnik, C. E. S., 
    A Residual Learning Approach for Unsteady Aerodynamic Load Prediction,
    \emph{International Forum on Aeroelasticity and Structural Dynamics}, Goëttingen, Germany, June 2026.
}
\officialurl{}

\usepackage{currfile-abspath}
\getmainfile 
\getabspath{\themainfile} 

\begin{document}

\maketitle


\begin{abstract}
This paper investigates the feasibility of using residual learning to improve unsteady aerodynamic load prediction for aeroelastic applications. The machine learning technique selected for the study is the long short-term memory (LSTM) neural network, which is used for its suitability for sequential data with aerodynamic memory effects. The approach is investigated for the NLR 7301 airfoil benchmark using high-fidelity CFD lift data for prescribed pitch and plunge motions in the transonic flow regime in the presence of
shock motion. An analytical unsteady aerodynamic model based on the Wagner function is used as a physics-based baseline, and the neural network is trained to learn the difference between the CFD lift coefficient and the Wagner prediction. The residual model is compared with a direct neural-network model trained to predict the CFD lift coefficient. The comparison includes feature and normalization studies, external benchmark cases, and leave-one-out and leave-family-out generalization tests across a range of sinusoidal and non-sinusoidal motions. The residual model performs best when its inputs align with the Wagner formulation variables, generally giving lower error and more consistent performance across training runs, though the direct model remains more accurate for some high-frequency cases. The residual model also generalizes better in the leave-one-out and leave-family-out tests, with a smaller increase in error than the direct model when entire motion families are withheld from training. Overall, the results indicate that residual learning shows promise as a modular approach for augmenting classical low-order aerodynamic theories, especially when the physics baseline removes a structured part of the aerodynamic response and leaves a lower-variance correction for the neural network to learn.
\end{abstract}

\section{Introduction}
\label{s1}

Accurate prediction of unsteady aerodynamic loads is essential for aeroelastic analysis, flutter assessment, and control-oriented modeling. The unsteady aerodynamic response depends not only on the instantaneous kinematics, but also on the reduced frequency and prior motion history. Classical thin-airfoil theories~\cite{theodorsen1949,bisplinghoff-ashley-1962}, such as Theodorsen's formulation and the Wagner function, decompose the attached-flow response into circulatory and noncirculatory contributions. These models provide physically interpretable low-order baselines, but they are derived under linear, incompressible, attached-flow assumptions. Their accuracy is therefore limited when compressibility effects, shock motion, viscous separation, dynamic-stall effects, or complex three-dimensional geometry become important. High-fidelity CFD can represent these effects more accurately, but at substantially higher computational cost, motivating reduced-order models that retain the relevant unsteady aerodynamic effects at a lower cost.

Neural-network-based ROMs approximate the input--output behavior of an aerodynamic system directly from data and avoid repeated solution of the governing equations during online prediction. For unsteady aerodynamic loads, this input--output relation is inherently sequential: the lift at a given time depends not only on the current pitch, plunge, flow conditions, and motion derivatives, but also on the prior motion history. Recurrent neural networks (RNNs) are therefore widely used for unsteady aerodynamic modeling because their hidden states provide a data-driven mechanism for carrying information from previous time steps into the current prediction. Prior studies have used recurrent multilayer perceptrons, radial-basis-function networks, neuro-fuzzy architectures, multi-kernel neural networks, and connected neural networks to predict aerodynamic coefficients, dynamic-stall loads, limit-cycle-oscillation behavior, and pitch--plunge responses at varying flow conditions~\cite{Faller1995,Suresh2003,Zhang2012,Winter2014,Mannarino2014,Ignatyev2015,KouMultiKernel2017,Winter2018}, with a comprehensive review provided by Kou and Zhang~\cite{Kou2021Review}. Other studies have combined neural networks with proper orthogonal decomposition (POD) or surrogate modeling techniques to predict distributed pressure or aerodynamic-field quantities rather than only integral coefficients~\cite{Park2013,Lindhorst2014,WinterPOD2016,Wang2020}. However, conventional gradient-trained RNNs can have difficulty learning long-term dependencies because gradients may decay over repeated time steps~\cite{Bengio1994}. Long short-term memory (LSTM) networks address this issue through gated cell states that retain sequence information over many time steps~\cite{Hochreiter1997}. Because LSTMs have been shown to preserve amplitude, phase, and history-dependent load behavior in unsteady aerodynamic ROMs~\cite{Wang2020}, they are adopted in the present work.

The NLR 7301 airfoil has served as a benchmark test case for data-driven unsteady aerodynamic modeling. Winter and Breitsamter used radial-basis-function neural networks to model unsteady aerodynamic loads for NLR 7301 pitch--plunge responses~\cite{Winter2014}. Wang et al. investigated techniques for improving neural-network aerodynamic ROMs for the same benchmark~\cite{Wang2019} and later developed multivariate recurrent models for both scalar lift histories and pressure-distribution predictions~\cite{Wang2020}. The Wang et al. dataset spans flow conditions from Mach 0.65 to 0.75 and includes single-harmonic, amplitude-modulated, Gaussian-pulse, and noise-perturbed pitch--plunge motion cases~\cite{Wang2020}. In those studies, the neural-network ROMs were trained to predict the full CFD lift coefficient directly, with the later multivariate RNN study also reconstructing pressure distributions~\cite{Wang2019,Wang2020}. As a result, the learned mapping is asked to represent quasi-steady lift, unsteady phase lag, wake memory, and compressibility-related trends within a single target. This combined target can be difficult to generalize when the test motion differs from the training motion family.
Moreover, the ROM remains purely data-driven: the physics is present only implicitly in the CFD training data, and no aerodynamic governing equation is explicitly retained in the learned model.

Physics-guided residual and layered learning strategies have been investigated to reduce the burden of learning the full high-fidelity response directly. Kou and Zhang introduced layered ROMs in which a fitted linear aerodynamic surrogate is corrected by a nonlinear neural-network model for nonlinear aerodynamic and aeroelastic prediction~\cite{Kou2017}. They later proposed a multi-fidelity framework in which an Euler-CFD baseline is corrected using data-driven residuals to approximate high-fidelity RANS responses for airfoils~\cite{Kou2019}. He and Schumann analyzed a physics-based deep recurrent residual neural-network model applied to fixed-wing flight-dynamics modeling~\cite{HeSchumann2020}. Other multi-fidelity neural-network approaches have fused low- and high-fidelity aerodynamic data for coefficient prediction~\cite{He2020} and aerodynamic shape optimization~\cite{Zhang2021}. Residual learning has also been used for aerodynamic coefficient prediction~\cite{Lee2023}, DeepONet-based multi-fidelity ROM correction~\cite{Demo2023}, and flight-load prediction with neural-network residual kriging~\cite{Yan2023}. More recent neural-network augmentation studies include increased-order modeling of aerodynamic nonlinearities~\cite{Feldwisch2025} and control-oriented aerodynamic-coefficient modeling for distributed-propulsion vehicles~\cite{Dong2025}.

In previous residual-learning studies, the baseline has typically been a fitted linear or state-space model, or a lower-fidelity CFD model. Whether an analytical low-fidelity model can provide an effective residual baseline for unsteady aerodynamic load prediction across a broad range of motion types remains insufficiently studied. The Wagner function provides a convenient physics-based baseline because it requires only the prescribed airfoil kinematics and flow speed, without additional fitting data or CFD solves, although its applicability is limited by its linear, incompressible and attached-flow assumptions. In addition, the robustness of physics-based residual learning under motion-type distribution shifts has not been systematically quantified, particularly when individual cases or entire motion families are withheld from training. The objective of this paper is therefore to quantify how a residual target based on an analytical Wagner indicial-function baseline changes the generalization behavior of an LSTM aerodynamic ROM relative to direct $c_l$ prediction for the NLR 7301 airfoil. The evaluation spans multiple pitch--plunge input families, including single-harmonic, amplitude-modulated harmonic, Gaussian-pulse, and noise-perturbed harmonic motions, using leave-one-out, leave-family-out, and external benchmark tests. By combining analytical unsteady aerodynamic theory with a data-driven residual correction, this study evaluates a practical methodology for improving unsteady aerodynamic predictions while preserving physical interpretability and computational efficiency for future aeroelastic, control-oriented, and multi-fidelity modeling applications.

The remainder of the paper is organized as follows. Section~\ref{s2} defines the NLR 7301 airfoil benchmark test case, training and testing data, and the Wagner baseline. Section~\ref{s3} presents the direct and residual model formulations, training and testing procedures, and error metrics used in the study. Section~\ref{s4} presents the results and Sec.~\ref{s5} summarizes the main conclusions and future work.

\section{Benchmark Problem and Physics Baseline}
\label{s2}

This section describes the aerodynamic test case, the training and external benchmark data, and the analytical baseline used throughout the study.

\subsection{NLR 7301 Airfoil CFD Data}
\label{s2-1}

\begin{figure}[htbp]
    \centering
    \begin{minipage}{0.245\linewidth}
        \centering
        \includegraphics[width=\textwidth]{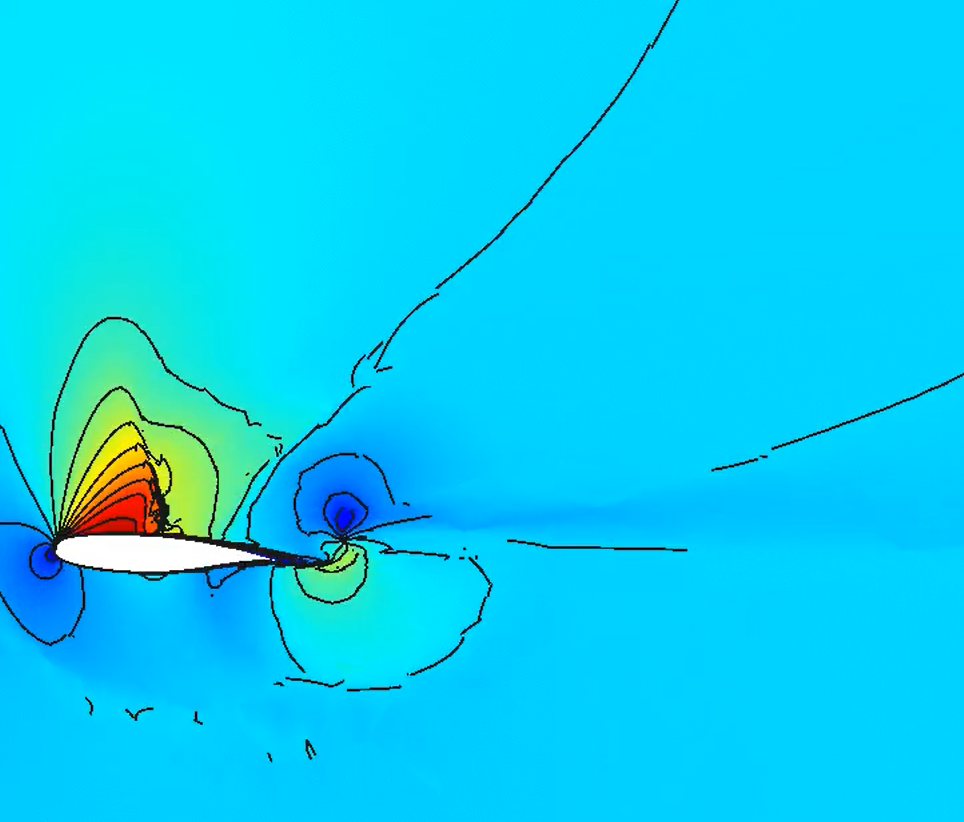}
    \end{minipage}
    \begin{minipage}{0.245\linewidth}
        \centering
        \includegraphics[width=\textwidth]{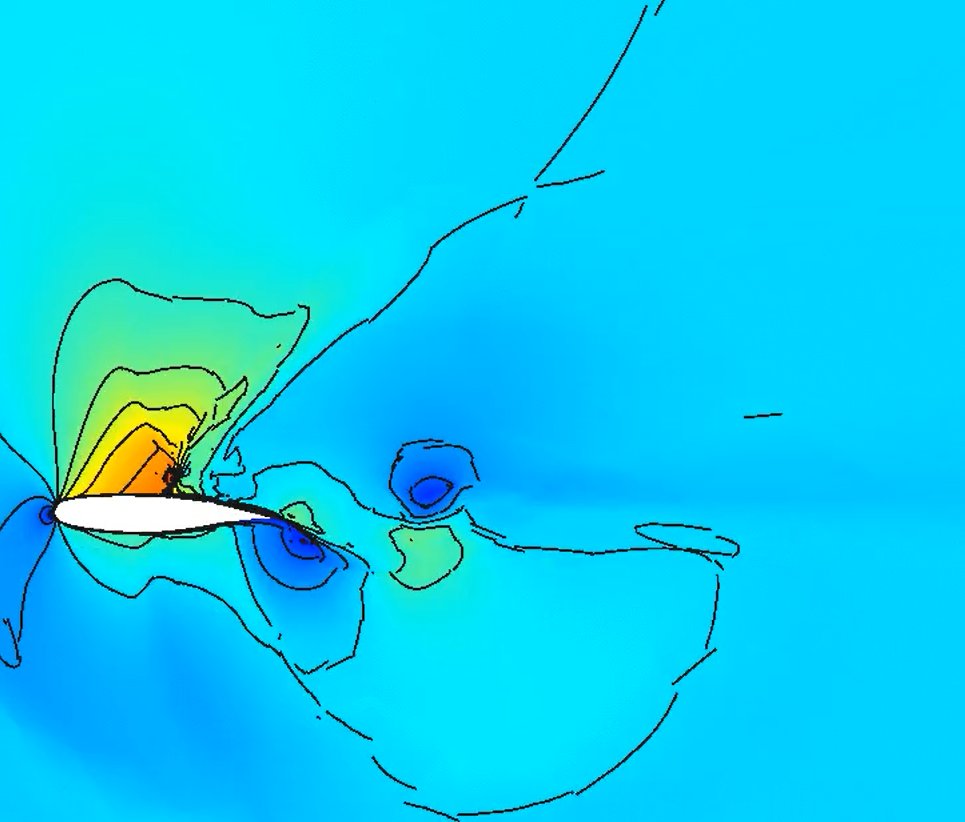}
    \end{minipage}
    \begin{minipage}{0.245\linewidth}
        \centering
        \includegraphics[width=\textwidth]{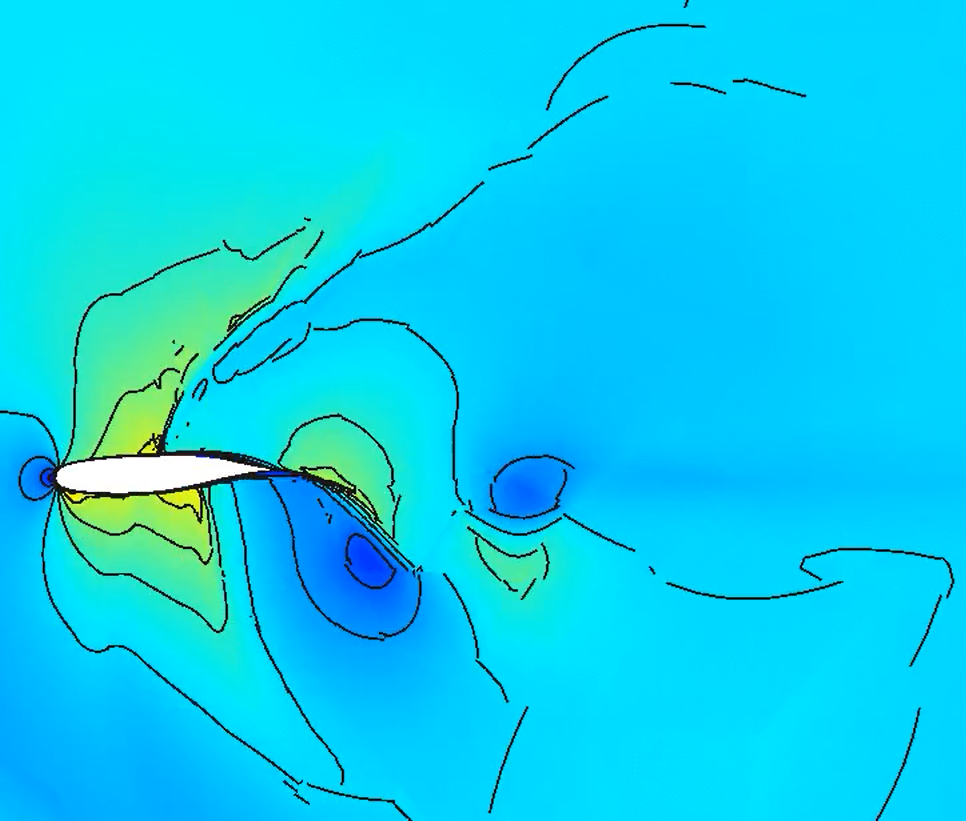}
    \end{minipage}
    \vspace{3mm}
    \caption{Snapshots of the NLR 7301 airfoil showing transonic shock motion from the Wang et al. high-fidelity CFD dataset~\cite{Wang2020}.}
    \label{fr:local_mach_snapshots}
\end{figure}

This work considers the two-dimensional NLR 7301 airfoil benchmark test case used in prior aerodynamic ROM studies~\cite{Winter2014,Wang2019,Wang2020}. The high-fidelity data used here are taken from the CFD database generated by Wang et al.~\cite{Wang2020} for a pitching and plunging airfoil using the in-house CFD code xflow~\cite{FidkowskiRoe2010}. The simulations were conducted at a Reynolds number of $2.1 \times 10^6$ over Mach numbers from $0.65$ to $0.75$. This Mach range includes transonic flow cases in which moving shocks appear over the airfoil (as shown in Fig.~\ref{fr:local_mach_snapshots}), increasing the complexity of the unsteady load prediction problem. The dataset used in this study comprises 48 CFD time histories for training and validation, together with three external benchmark time histories for testing.

The available input histories include the prescribed pitch angle $\alpha(t)$, plunge displacement $h(t)$, Mach number $M$, and the corresponding derivatives $\dot{\alpha}(t)$ and $\dot{h}(t)$. The pitch and plunge motion inputs are analytically defined, and the first-order derivatives are provided directly with the CFD data. Second-order derivatives, $\ddot{\alpha}(t)$ and $\ddot{h}(t)$, when required as model inputs, are estimated from the recorded position histories using second-order central finite differences.
The available aerodynamic outputs include integral and distributed aerodynamic quantities, but the present study focuses only on lift coefficient prediction, $c_l(t)$.

The reduced frequency used throughout this paper is defined as
\begin{equation}
    k = \frac{2\omega b}{U} ,
    \label{eq:reduced_frequency}
\end{equation}
where $\omega$ is the angular oscillation frequency, $b$ is the semi-chord, $2b$ is the full chord, and $U$ is the free-stream speed. This convention is consistent with the study in~\cite{Wang2020}. 
The time coordinate in the raw CFD files is the nondimensional time, defined as
\begin{equation}
    \tau = t_\mathrm{xflow} = \frac{U t_\mathrm{phys}}{2b},
    \label{eq:xflow_time}
\end{equation}
where $t_\mathrm{phys}$ is physical time. For this dataset, the semi-chord is $b=0.09$\,m (full chord $2b=0.18$\,m), and $c=340.294$\,m/s is the speed of sound used for dimensional conversion. If dimensional time is needed for interpretation, $t_\mathrm{phys}=2b\tau/U=2b\tau/(M c)$. All training, validation, and external testing are performed on the original $\tau$-indexed solver samples, without conversion to physical seconds or resampling to a common time grid. The nondimensional time step varies across motion cases, and all data splits are made by complete motion case to avoid leakage of aerodynamic history information across subsets.

\subsection{Training and External Benchmark Cases}
\label{s2-2}

The training dataset contains 48 cases spanning four prescribed-motion families, summarized in Table~\ref{t:motion_families}. The single-harmonic cases provide controlled frequency and Mach variation over a grid of $k$ values, while the three non-sinusoidal families introduce amplitude modulation, transient pulse behavior, and noise-perturbed harmonic samples~\cite{Wang2020}.

\begin{figure}
    \centering
    \subfigure[Excitation]{%
        \begin{minipage}{0.48\linewidth}
            \centering
            \includegraphics[width=\textwidth]{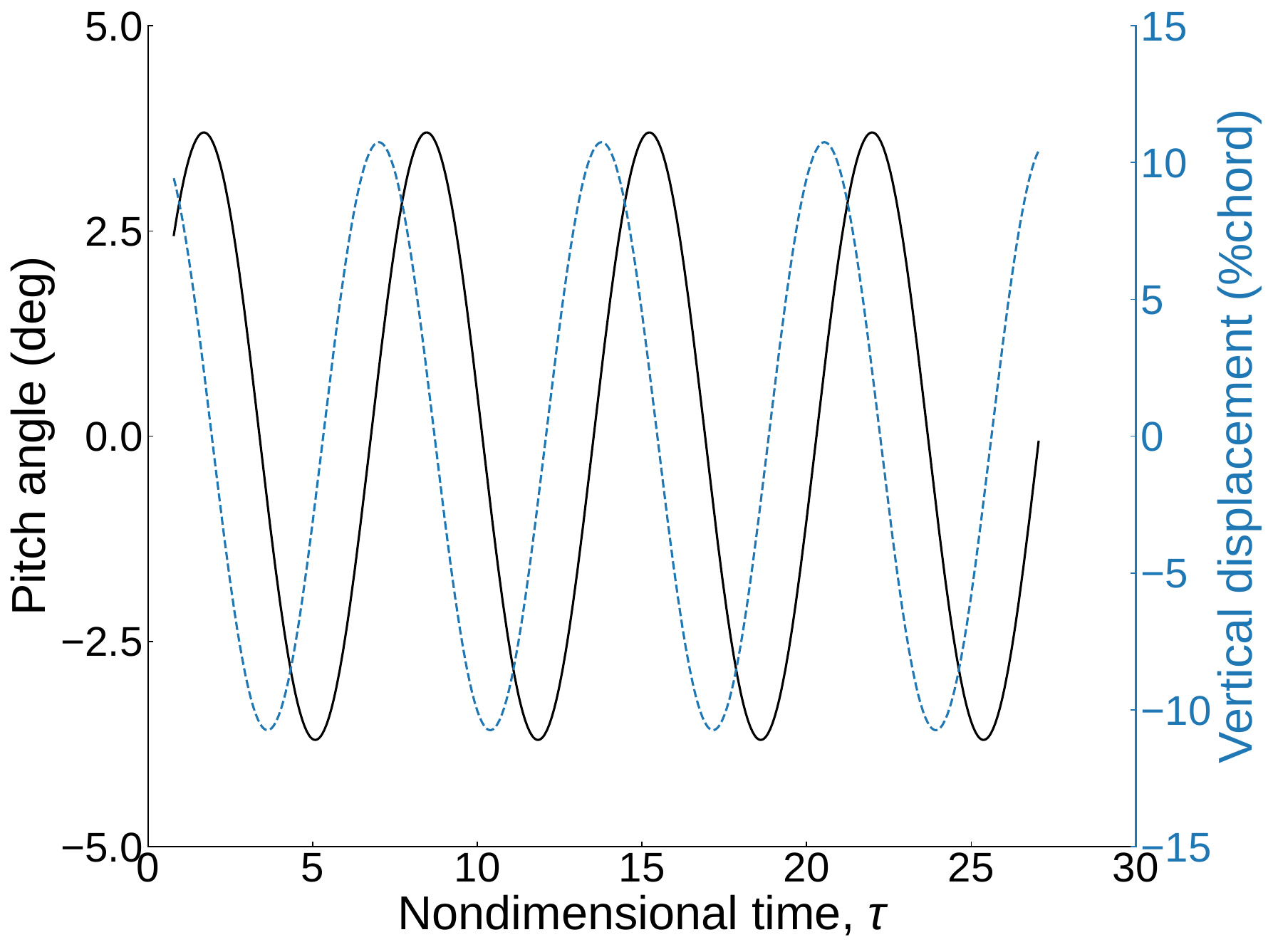}
        \end{minipage}%
    }\hspace{2mm}
    \subfigure[Lift coefficient]{%
        \begin{minipage}{0.48\linewidth}
            \centering
            \includegraphics[width=\textwidth]{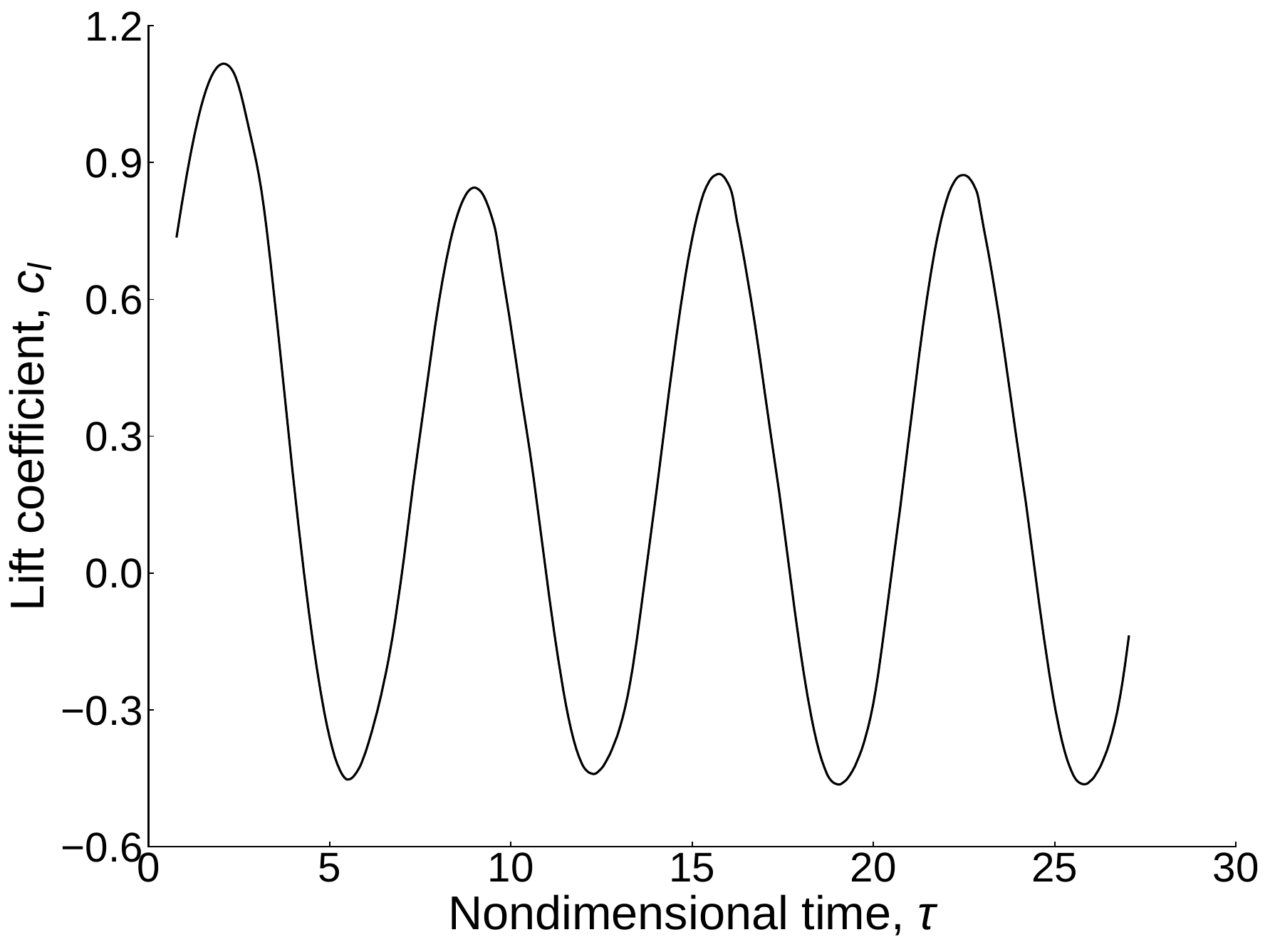}
        \end{minipage}%
    }
    \vspace{2mm}
    \caption{Representative single-harmonic training case.}
    \label{fr:training_sh}
\end{figure}

\begin{figure}
    \centering
    \subfigure[Excitation]{%
        \begin{minipage}{0.48\linewidth}
            \centering
            \includegraphics[width=\textwidth]{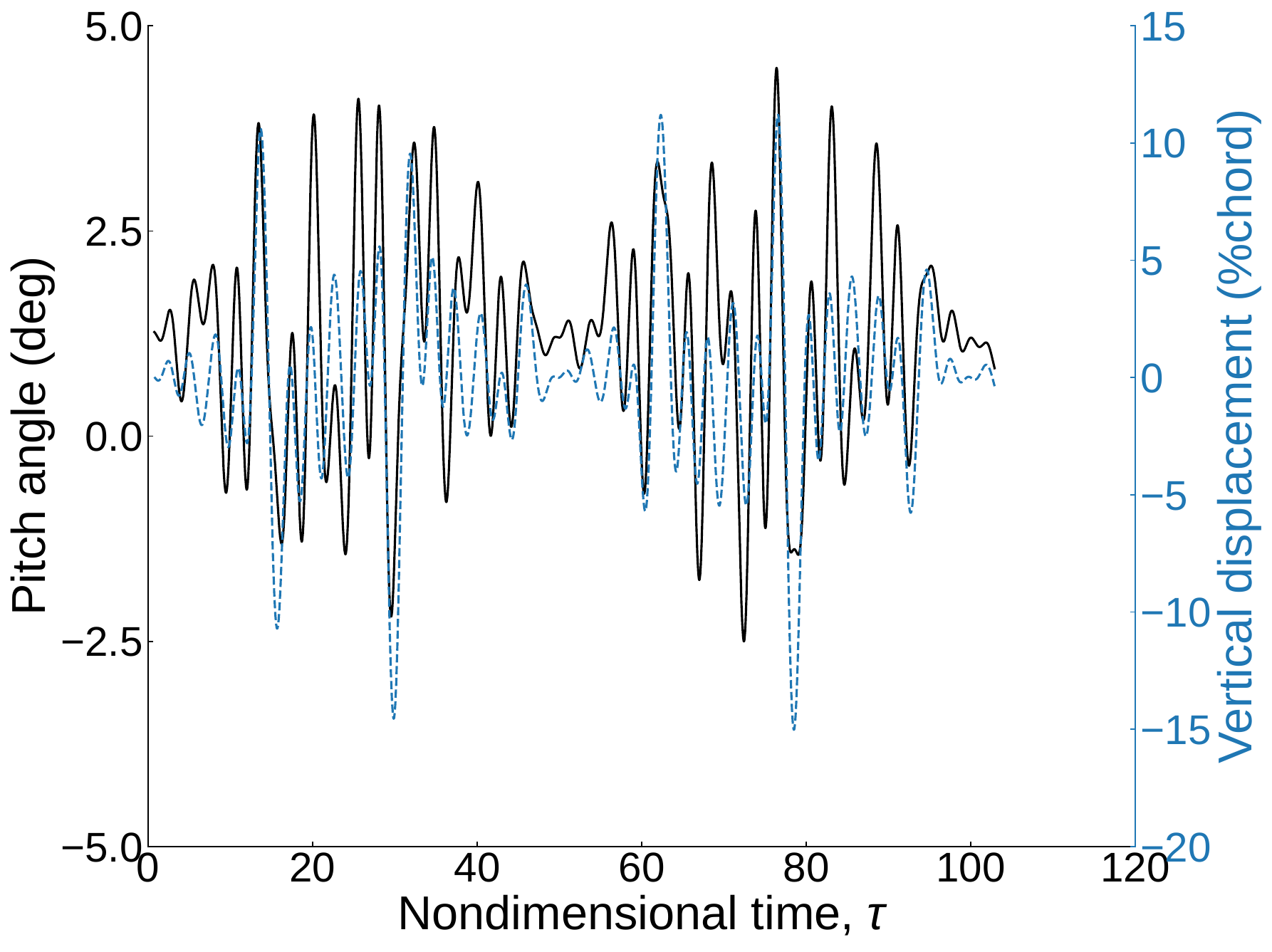}
        \end{minipage}%
    }\hspace{2mm}
    \subfigure[Lift coefficient]{%
        \begin{minipage}{0.48\linewidth}
            \centering
            \includegraphics[width=\textwidth]{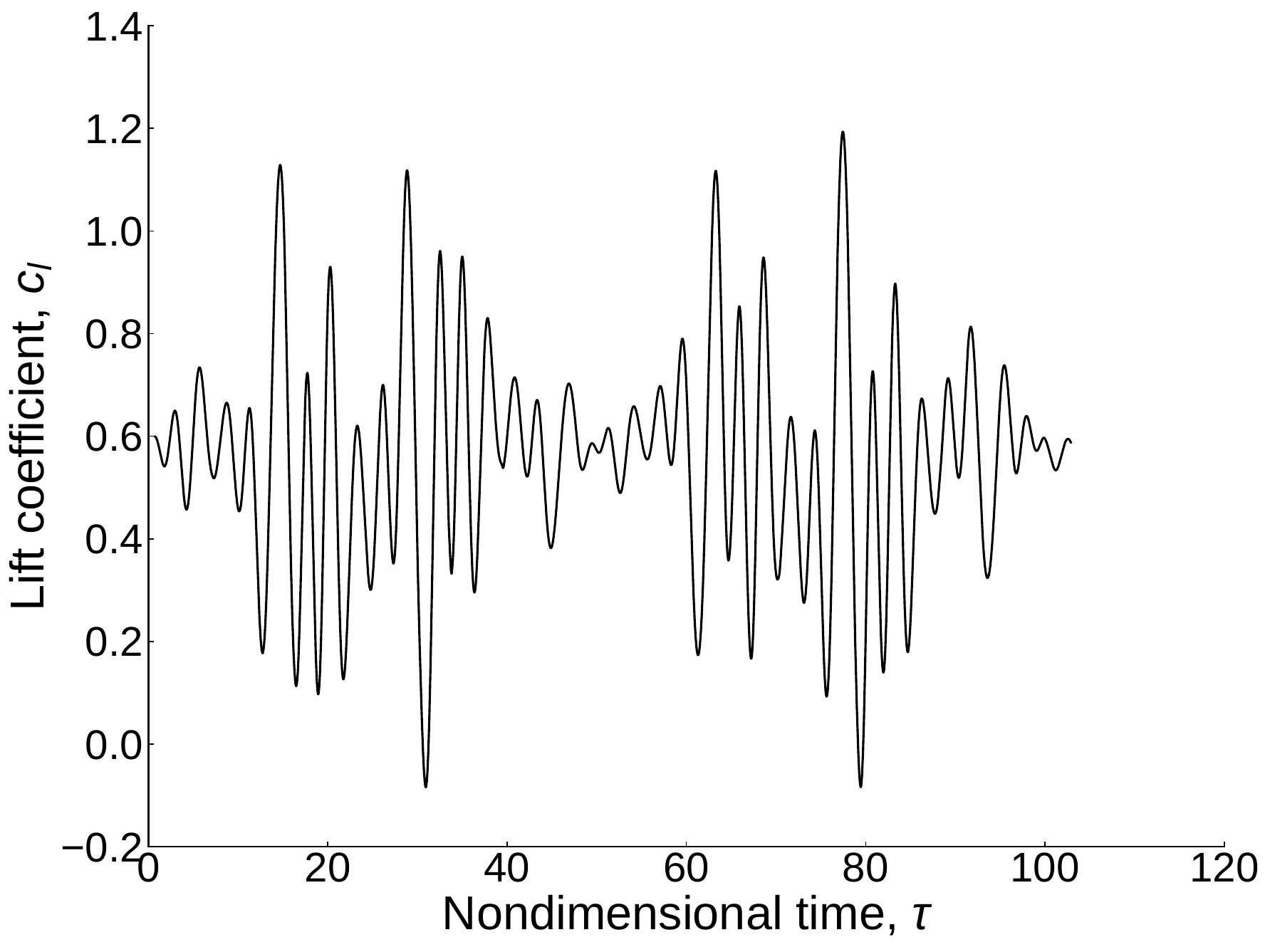}
        \end{minipage}%
    }
    \vspace{2mm}
    \caption{Representative amplitude-modulated harmonic training case.}
    \label{fr:training_am}
\end{figure}

\begin{figure}
    \centering
    \subfigure[Excitation]{%
        \begin{minipage}{0.48\linewidth}
            \centering
            \includegraphics[width=\textwidth]{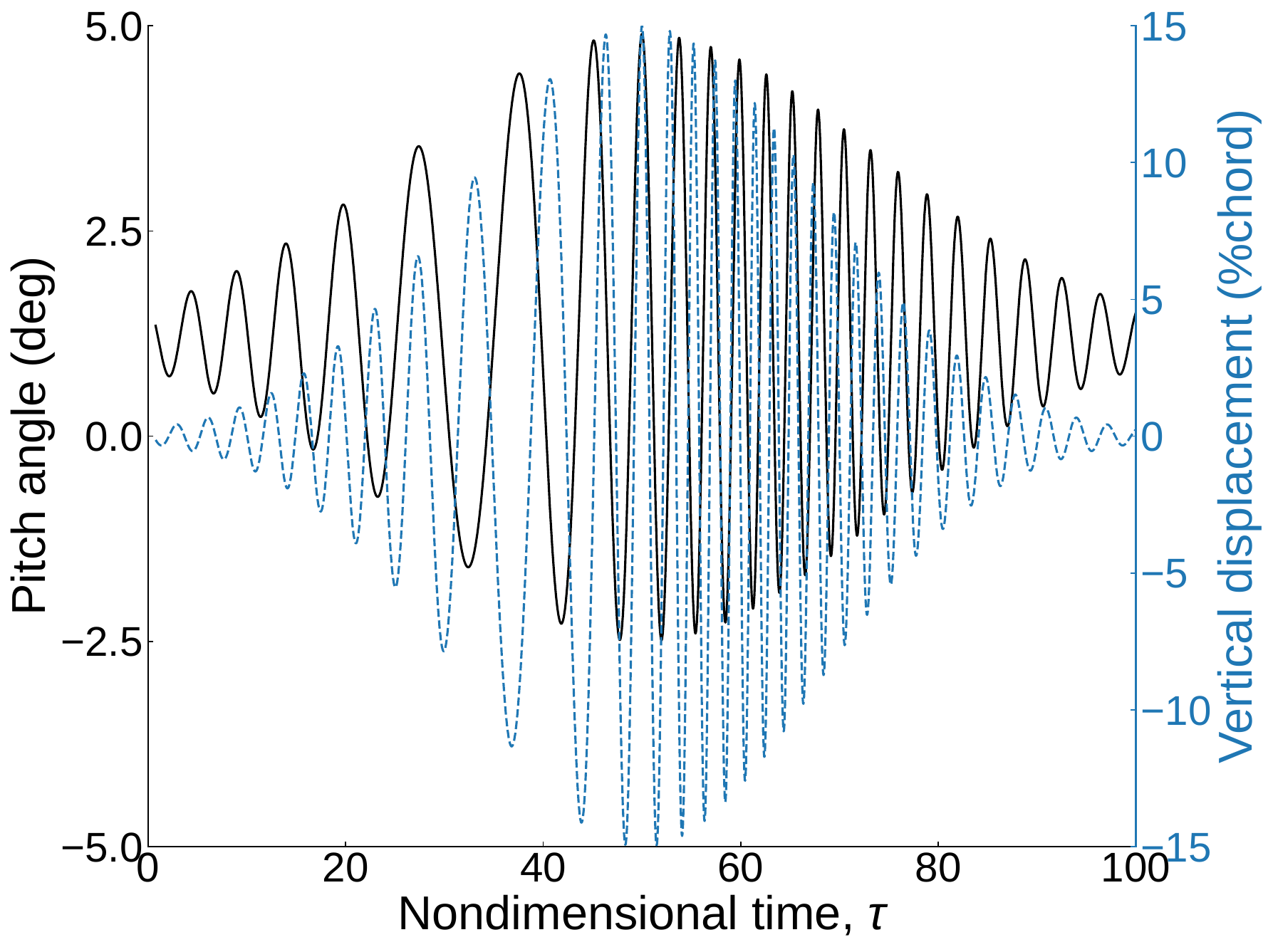}
        \end{minipage}%
    }\hspace{2mm}
    \subfigure[Lift coefficient]{%
        \begin{minipage}{0.48\linewidth}
            \centering
            \includegraphics[width=\textwidth]{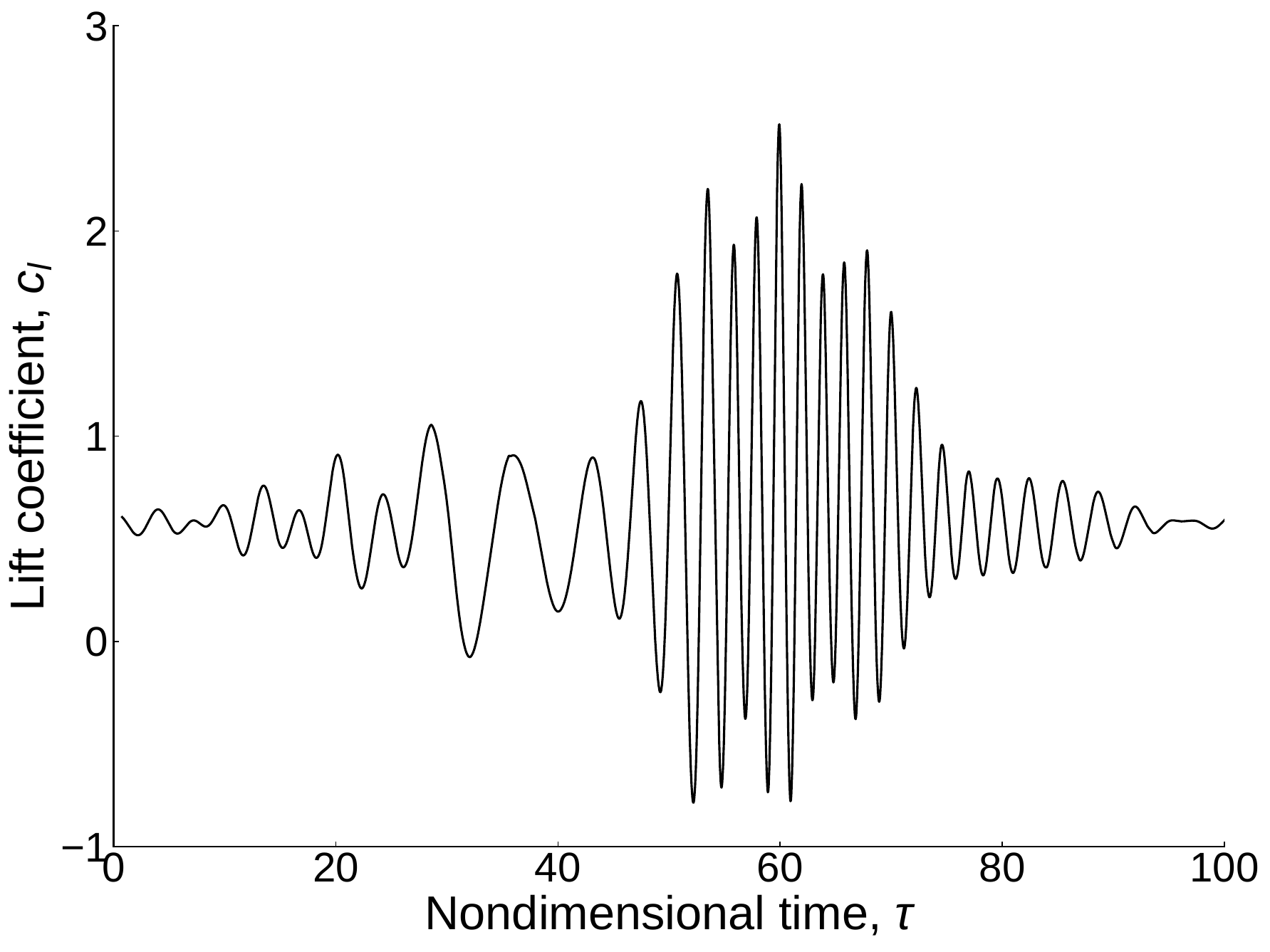}
        \end{minipage}%
    }
     \vspace{2mm}
    \caption{Representative Gaussian-pulse training case.}
    \label{fr:training_gp}
\end{figure}

\begin{figure}
    \centering
    \subfigure[Excitation]{%
        \begin{minipage}{0.48\linewidth}
            \centering
            \includegraphics[width=\textwidth]{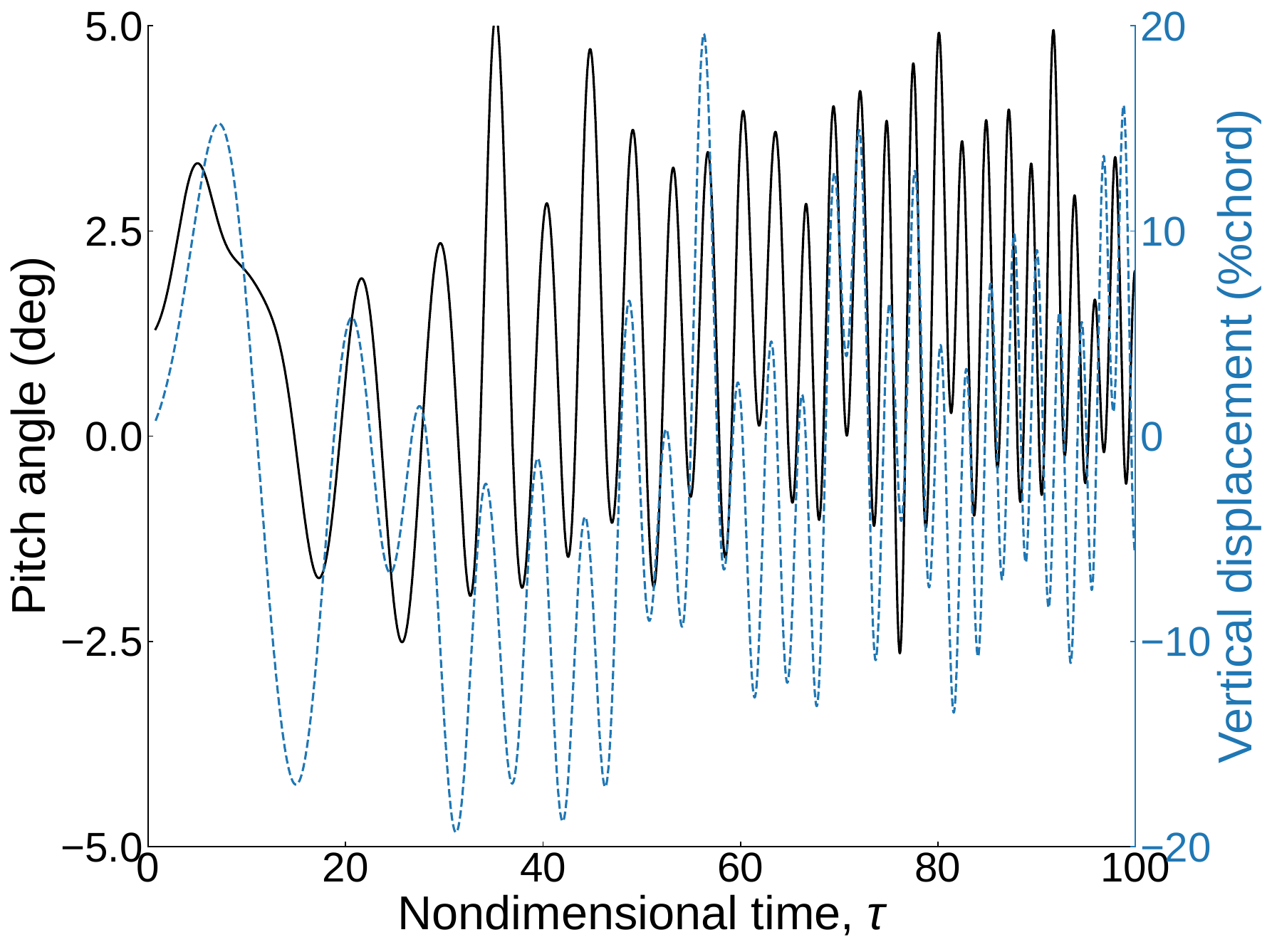}
        \end{minipage}%
    }\hspace{2mm}
    \subfigure[Lift coefficient]{%
        \begin{minipage}{0.48\linewidth}
            \centering
            \includegraphics[width=\textwidth]{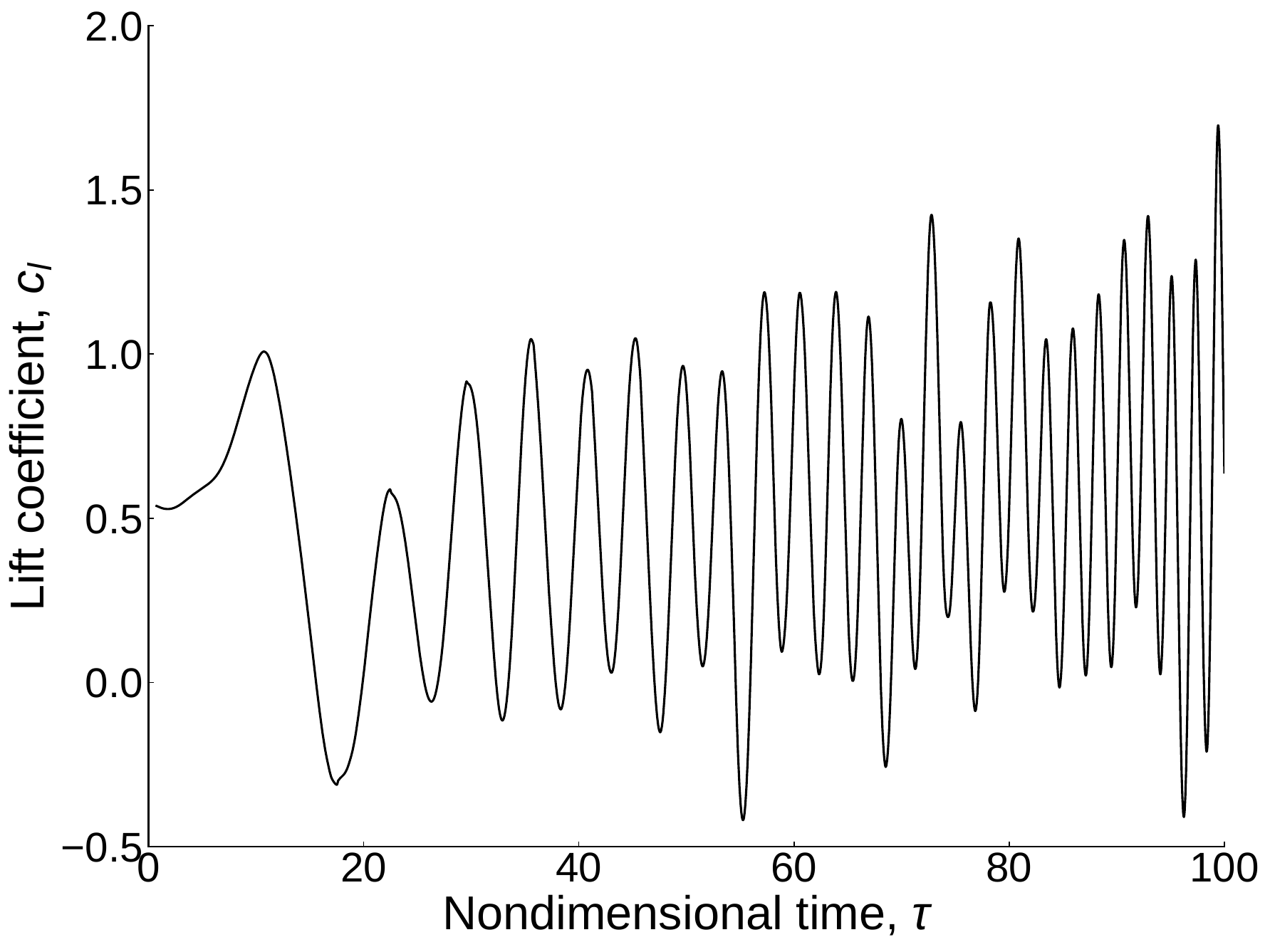}
        \end{minipage}%
    }
     \vspace{2mm}
    \caption{Representative noise-perturbed harmonic training case.}
    \label{fr:training_rl1}
\end{figure}

\begin{table}
    \centering
    \caption{Training motion families in the high-fidelity CFD dataset.}
    \vspace{3mm}
    \begingroup
    \setlength{\tabcolsep}{4pt}
    \renewcommand{\arraystretch}{1.15}
    \begin{tabularx}{\textwidth}{@{}>{\raggedright\arraybackslash}p{0.22\textwidth}>{\centering\arraybackslash}p{0.09\textwidth}>{\raggedright\arraybackslash}X>{\raggedright\arraybackslash}X@{}}
    \toprule
    Family & Cases & Mach numbers & Frequency content \\
    \midrule
    Single harmonic & 24 & 0.65, 0.70, 0.75 & $k\in[0.088,3.007]$ \\
    Amplitude-modulated harmonic & 6 & 0.65, 0.67, 0.69, 0.71, 0.73, 0.75 & Components about $k\approx2.16$ \\
    Gaussian pulse & 6 & 0.65, 0.67, 0.69, 0.71, 0.73, 0.75 & Approx. $k\in[0.63,3.02]$ \\
    Noise-perturbed harmonic & 12 & 0.65, 0.67, 0.69, 0.71, 0.73, 0.75 & Approx. $k\in[0.03, 3.65]$ \\
    \bottomrule
    \end{tabularx}
    \endgroup
    \label{t:motion_families}
\end{table}
The physical character of each family determines its role in the generalization studies. Representative examples of each family are shown in Figs.~\ref{fr:training_sh}--\ref{fr:training_rl1}. The single-harmonic cases are narrow-spectrum sinusoidal inputs in pitch and plunge at fixed frequency and Mach, and they serve as the primary family for leave-one-out and leave-family-out comparisons. The amplitude-modulated harmonic cases impose a low-frequency amplitude envelope onto a sinusoidal carrier near $k \approx 2.16$. The resulting pitch and plunge histories have a semi-periodic behavior in which the instantaneous amplitude varies slowly relative to the oscillation period, producing a narrowband but non-stationary excitation. The Gaussian pulse cases concentrate excitation energy in the middle of the simulation window, with pitch and plunge driven by a chirp-like modulated pulse. The transient ramp-up and ramp-down phases require the aerodynamic model to capture initial transient and decay behavior that sinusoidal training data do not represent. The noise-perturbed harmonic cases are constructed by superimposing noise onto a single-harmonic base, yielding non-periodic inputs whose frequency content spans the full $k$ range used in training. For each Mach number, the two variants use the same pitch input but opposite sign for the plunge inputs, so $\alpha_2(t)=\alpha_1(t)$ and $h_2(t)=-h_1(t)$.

\begin{figure}
    \centering
    \subfigure[Excitation]{%
        \begin{minipage}{0.48\linewidth}
            \centering
            \includegraphics[width=\textwidth]{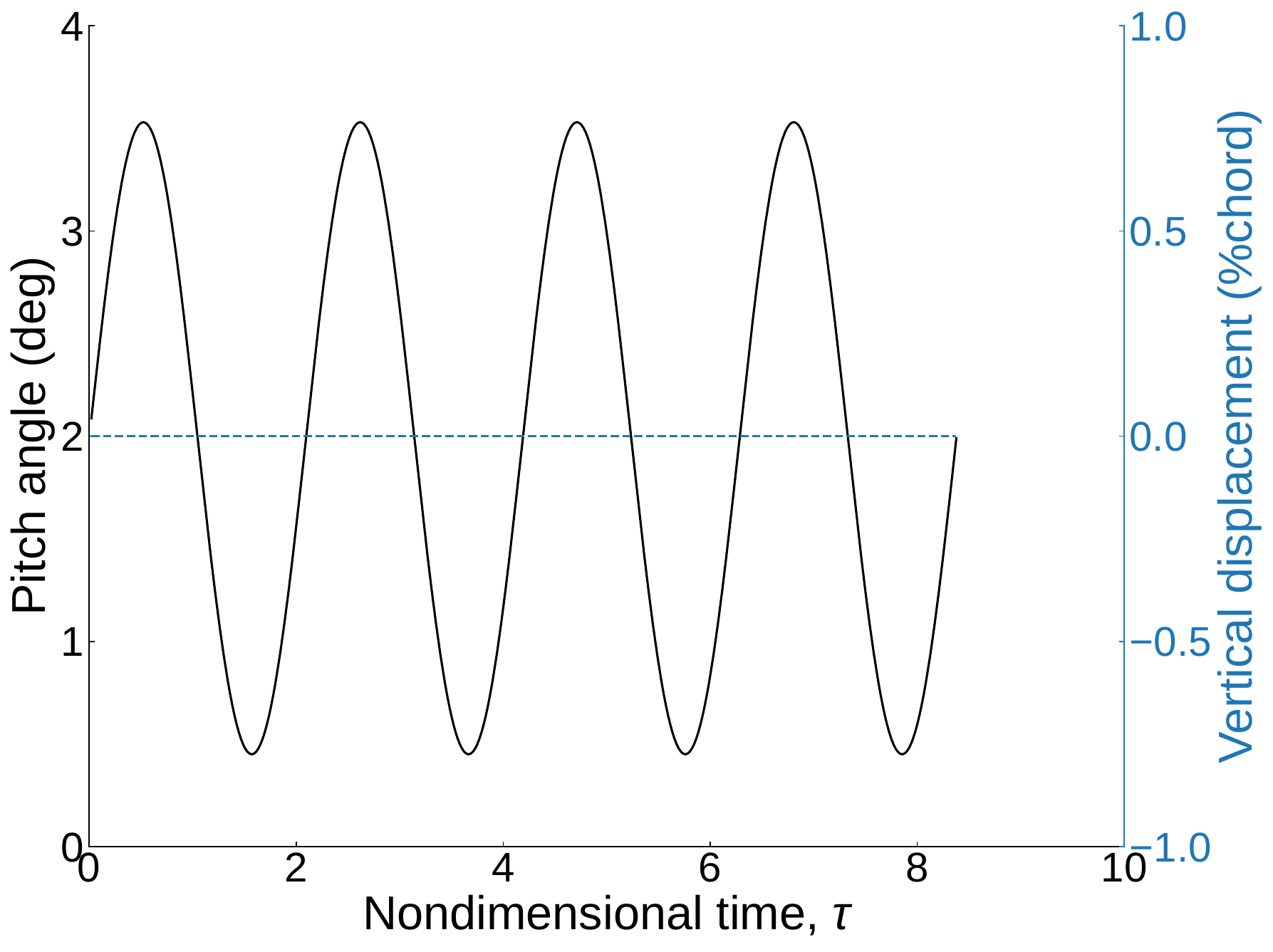}
        \end{minipage}%
    }\hspace{2mm}
    \subfigure[Lift coefficient]{%
        \begin{minipage}{0.48\linewidth}
            \centering
            \includegraphics[width=\textwidth]{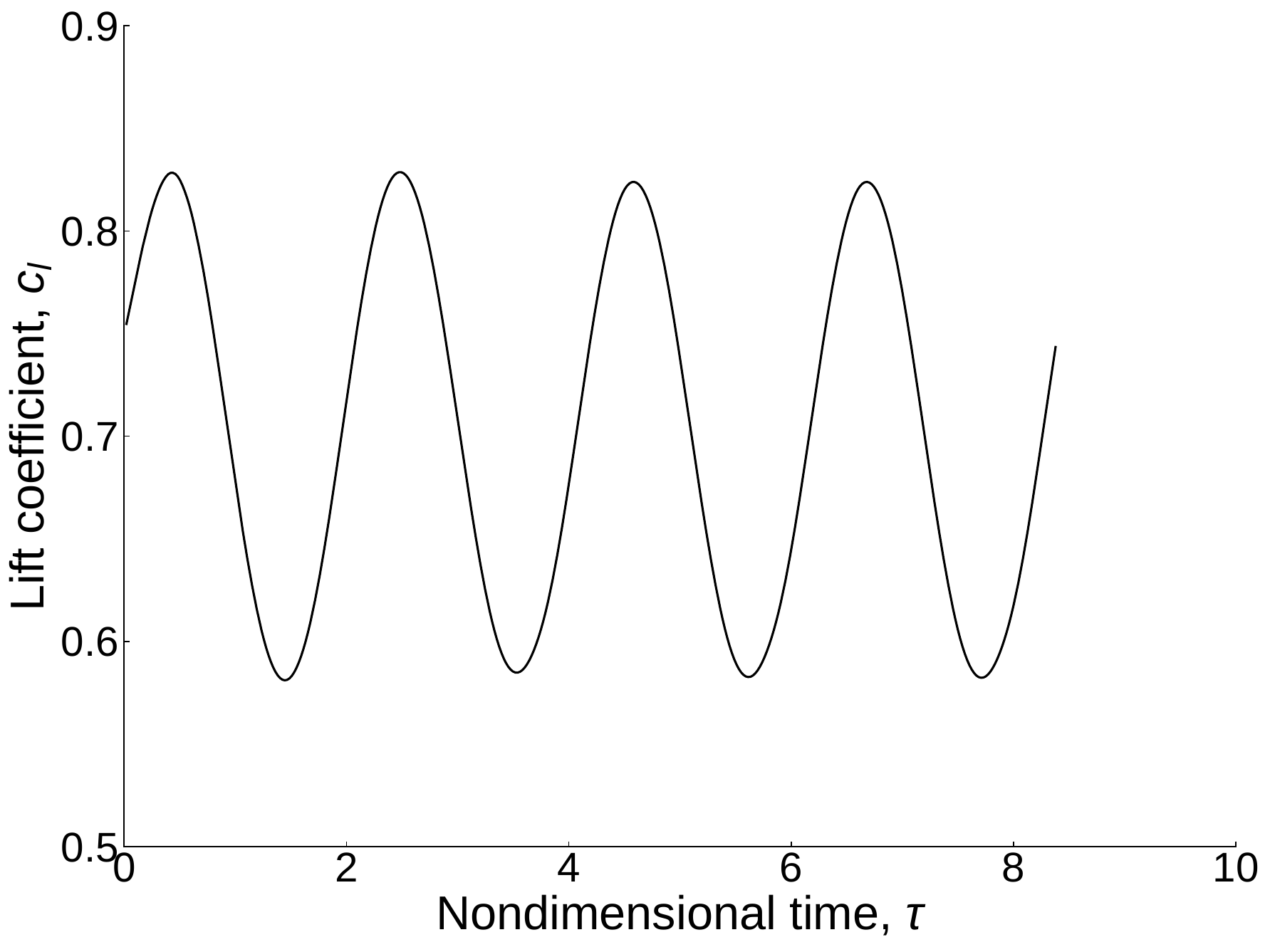}
        \end{minipage}%
    }
     \vspace{2mm}
    \caption{Pitch-only external benchmark case.}
    \label{fr:test_case02}
\end{figure}

\begin{figure}
    \centering
    \subfigure[Excitation]{%
        \begin{minipage}{0.48\linewidth}
            \centering
            \includegraphics[width=\textwidth]{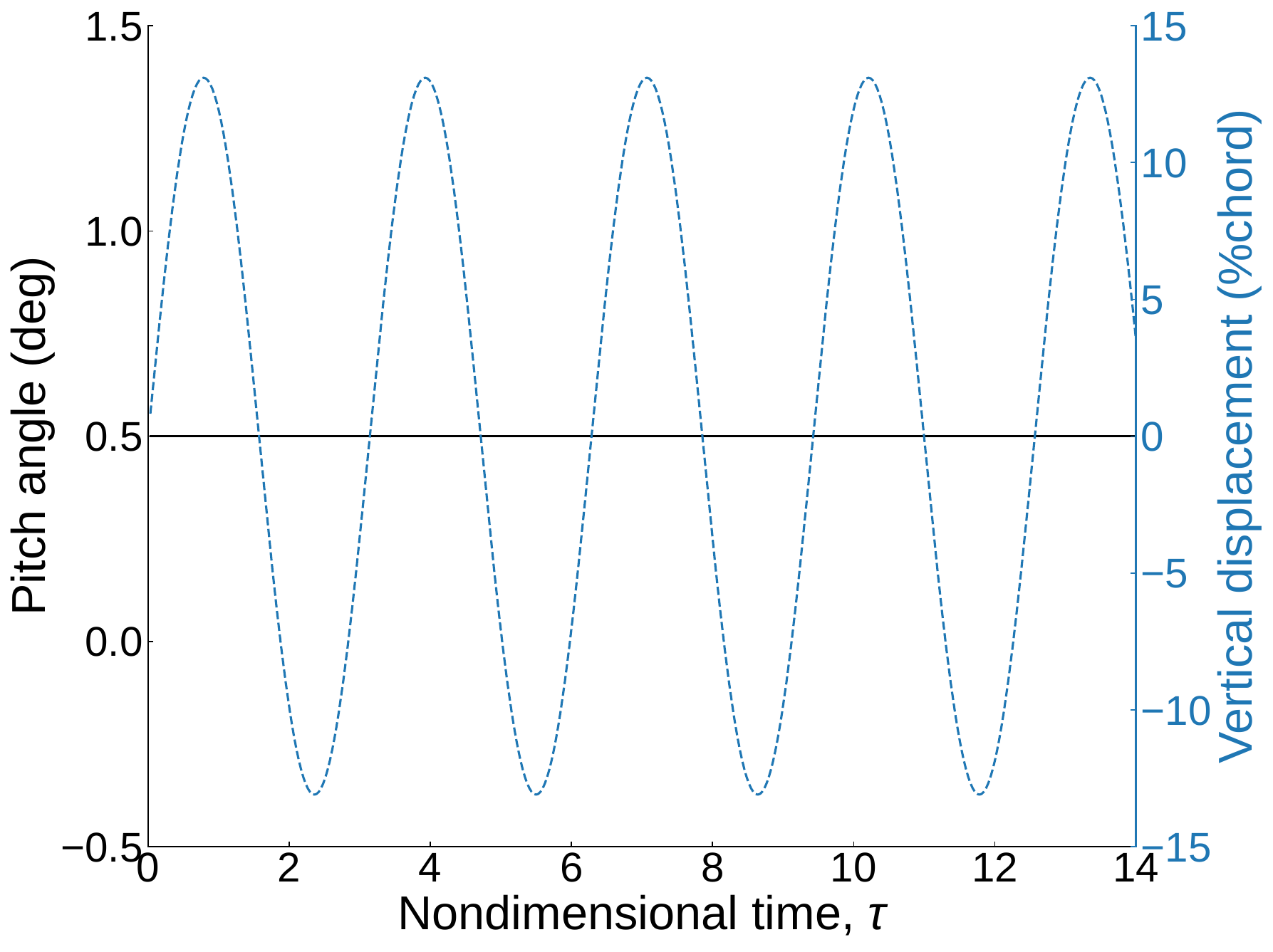}
        \end{minipage}%
    }\hspace{2mm}
    \subfigure[Lift coefficient]{%
        \begin{minipage}{0.48\linewidth}
            \centering
            \includegraphics[width=\textwidth]{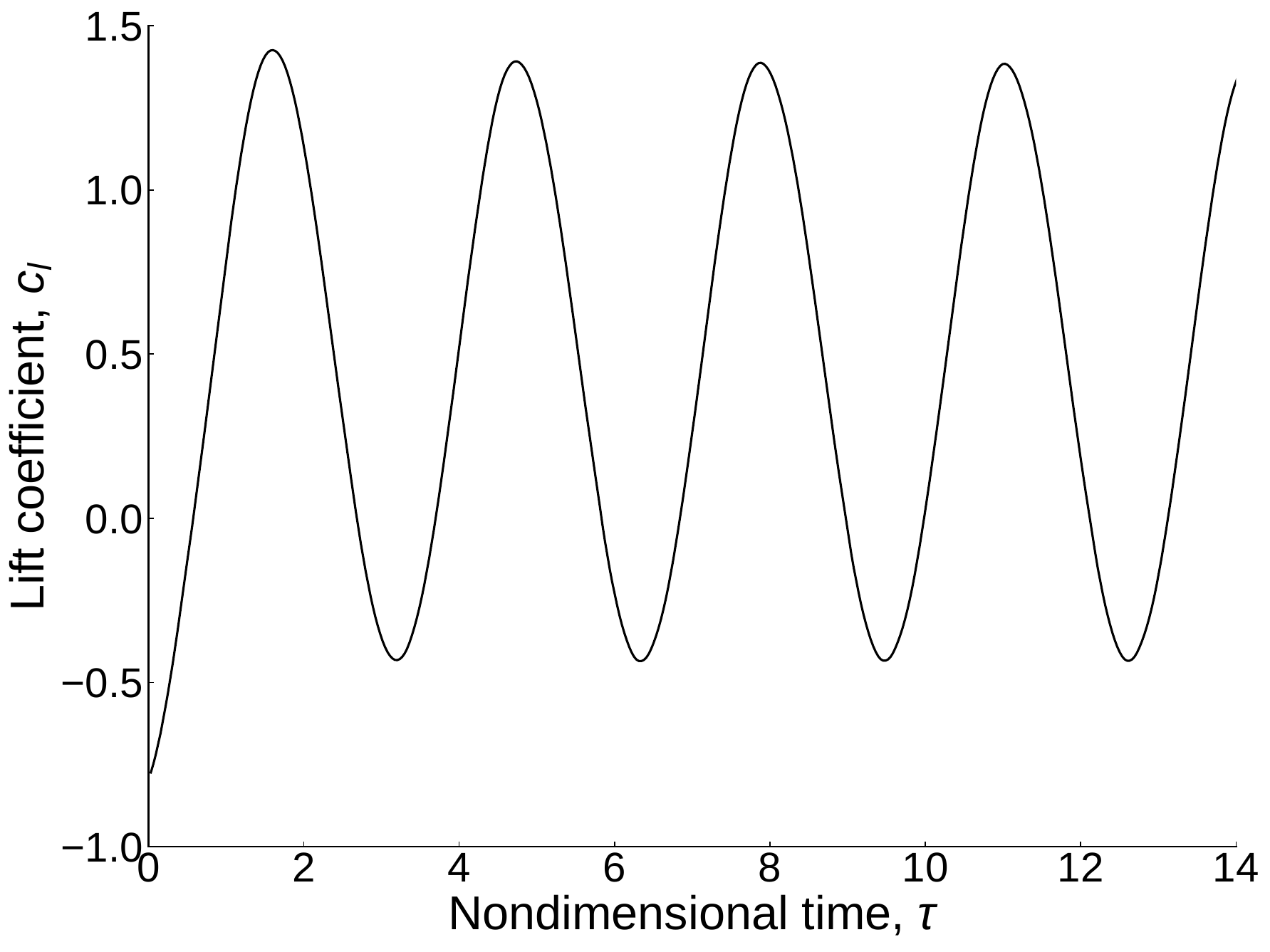}
        \end{minipage}%
    }
     \vspace{2mm}
    \caption{Plunge-only external benchmark case.}
    \label{fr:test_case10}
\end{figure}

\begin{figure}
    \centering
    \subfigure[Excitation]{%
        \begin{minipage}{0.48\linewidth}
            \centering
            \includegraphics[width=\textwidth]{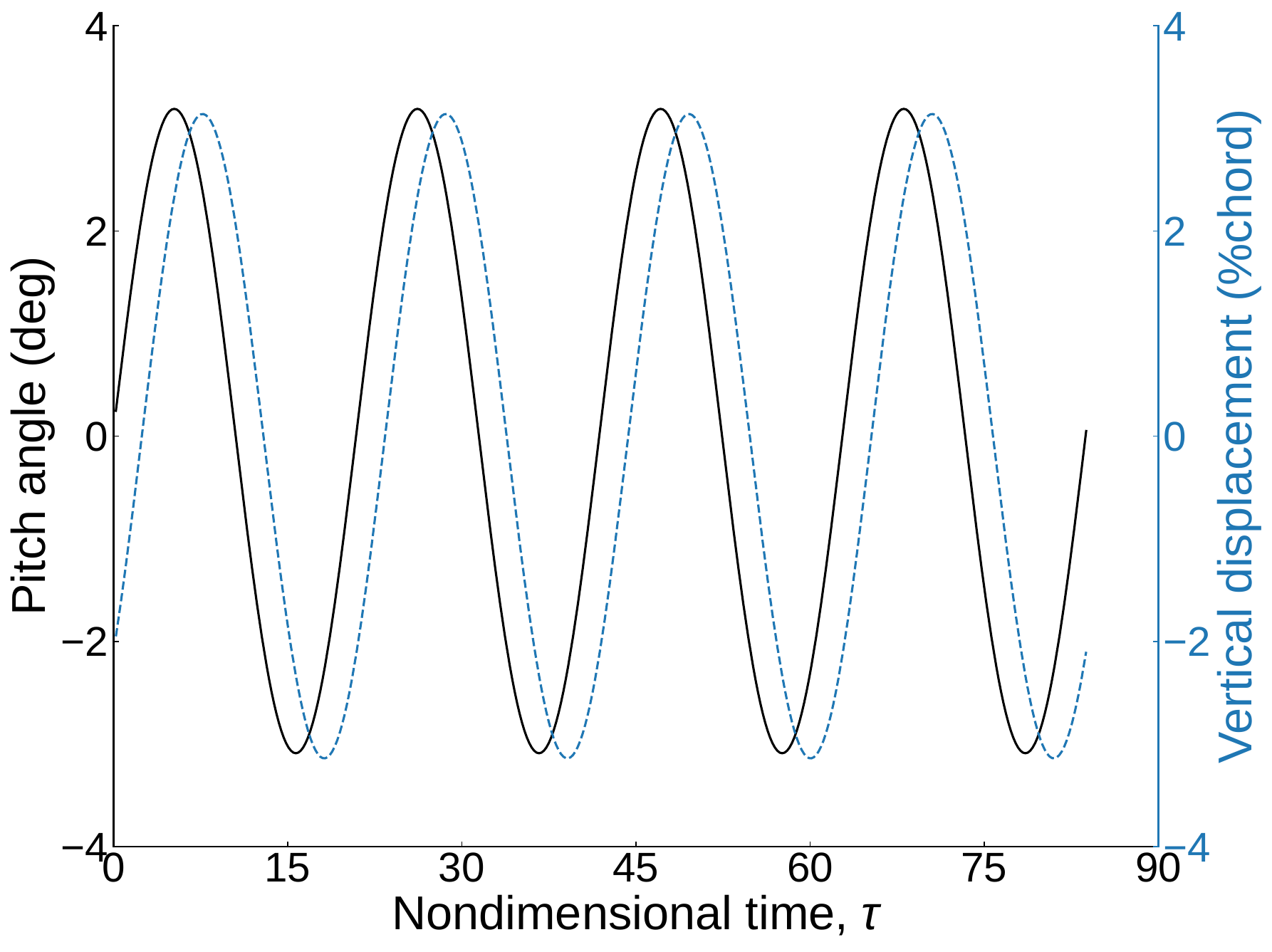}
        \end{minipage}%
    }\hspace{2mm}
    \subfigure[Lift coefficient]{%
        \begin{minipage}{0.48\linewidth}
            \centering
            \includegraphics[width=\textwidth]{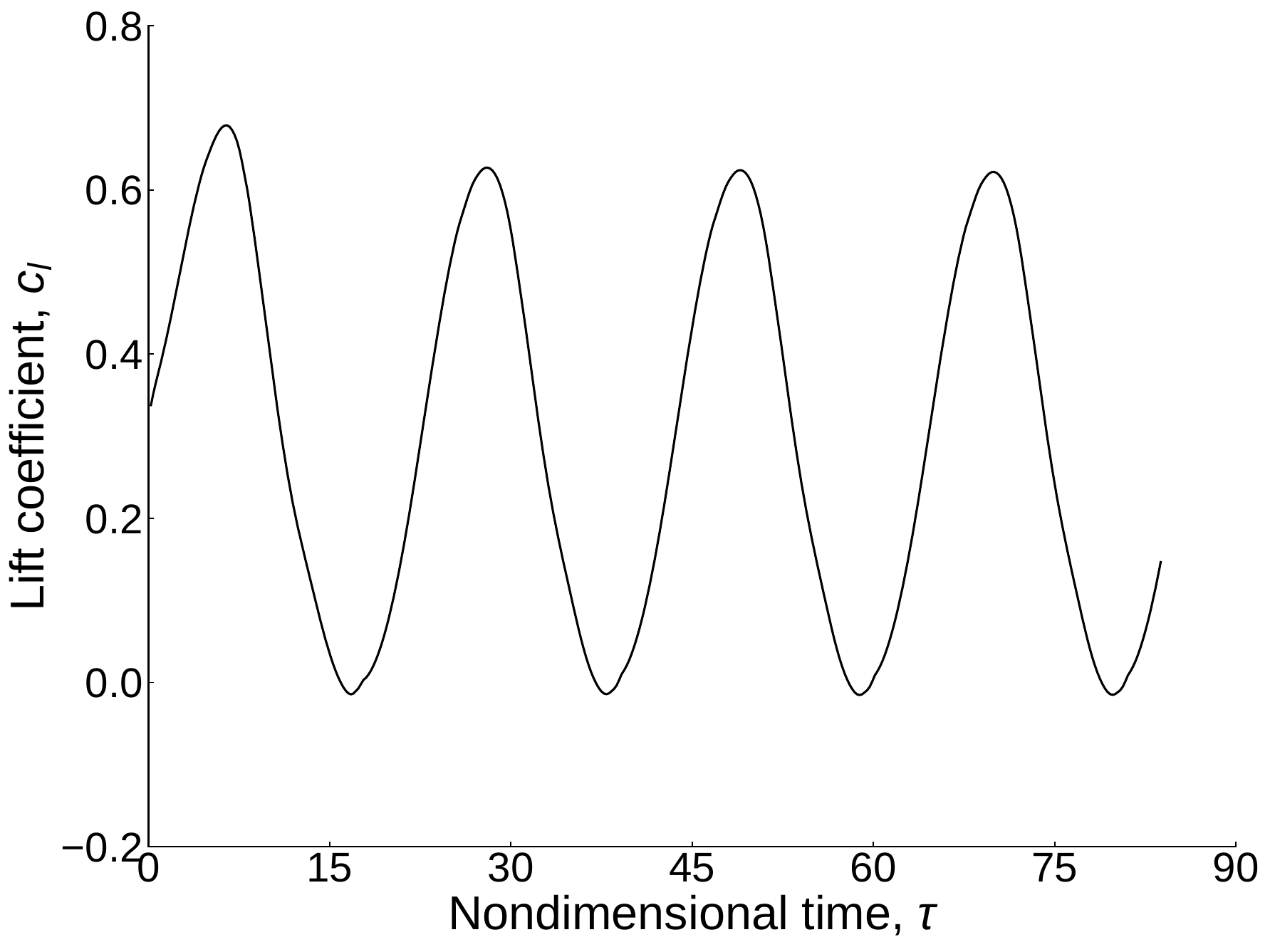}
        \end{minipage}%
    }
     \vspace{2mm}
    \caption{Coupled pitch--plunge external benchmark case.}
    \label{fr:test_case15}
\end{figure}

In addition to the training data, three external benchmark cases are used as the verification set. Table~\ref{t:external_cases} gives the Mach number and reduced frequency for all three cases, and their prescribed kinematics and CFD lift-coefficient time histories are shown in Figs.~\ref{fr:test_case02}--\ref{fr:test_case15}. These benchmark cases are not used for fitting and are retained as complete time histories in the same $\tau=t_\mathrm{xflow}$ convention as the training cases. The three cases cover distinct kinematic regimes: pure pitch at high reduced frequency, pure plunge at moderate reduced frequency, and coupled pitch--plunge motion at low reduced frequency. 
\begin{table}
    \centering
    \caption{External benchmark cases used for testing.}
    \vspace{3mm}
    \begin{tabular}{lrr}
    \toprule
    Case & $M$ & $k$ \\
    \midrule
    Pitch only & 0.705 & 3.01 \\
    Plunge only & 0.720 & 2.00 \\
    Coupled pitch--plunge & 0.750 & 0.30 \\
    \bottomrule
    \end{tabular}
    \label{t:external_cases}
\end{table}

\subsection{Wagner Baseline}
\label{s2-4}

This subsection defines the analytical aerodynamic baseline used in the residual learning framework. The Wagner function was selected because it provides a physics-based unsteady lift prediction from the prescribed pitch--plunge kinematics and free-stream conditions alone, without requiring additional CFD computations. It decomposes the unsteady lift for a thin airfoil undergoing pitch and
plunge motion in incompressible flow into a noncirculatory added mass term and a circulatory term obtained from a Duhamel convolution. The Wagner indicial response equations are written in terms of the nondimensional time
\begin{equation}
    s = \frac{U t_\mathrm{phys}}{b},
    \label{eq:nondim_time}
\end{equation}
where $U$ is the free-stream speed and $b$ is the airfoil semi-chord. Since the CFD histories are stored in the full-chord time coordinate of Eq.~(\ref{eq:xflow_time}), the Wagner function is evaluated using
\begin{equation}
    s = 2\tau .
    \label{eq:wagner_time_relation}
\end{equation}
The CFD plunge displacement is positive upward while the Wagner formulation uses the thin-airfoil sign convention, so the CFD data plunge variable is reversed in sign before evaluating the baseline. The effective downwash at the three-quarter-chord point is
\begin{equation}
    w_{3/4} = -\!\left[\dot{h} + U\alpha + b\!\left(\tfrac{1}{2} - a\right)\dot{\alpha}\right],
    \label{eq:downwash}
\end{equation}
where $h$ denotes the plunge displacement and $\alpha$ is the pitch angle. Using the above, the lift is given by
\begin{align}
    L &= \underbrace{\pi\rho b^2\!\left(\ddot{h} + U\dot{\alpha} - ba\ddot{\alpha}\right)}_{\text{noncirculatory}} \nonumber \\
      &\quad - \rho U b \underbrace{\left[2\pi w_{3/4}(s)\,\phi(0)
        + 2\pi\!\int_0^s \frac{\mathrm{d}\phi}{\mathrm{d}\sigma}\,w_{3/4}(s-\sigma)\,\mathrm{d}\sigma\right]}_{\text{circulatory}},
    \label{eq:wagner_lift}
\end{align}
where $\rho$ is the air density and $\phi$ is the Wagner indicial function, given by 
\begin{equation}
    \phi(\sigma) = 1 - 0.203\,e^{-0.072\sigma} - 0.236\,e^{-0.261\sigma} - 0.06\,e^{-0.8\sigma},
    \label{eq:wagner_phi}
\end{equation}
The lift coefficient is normalized by the dynamic pressure and full chord,
\begin{equation}
    c_l = \frac{L}{\tfrac{1}{2}\rho U^2(2b)}.
    \label{eq:cl_norm}
\end{equation}
In this study, three variants of this analytical baseline are defined and compared to select the most suitable physics-based baseline for residual learning: quasi-steady, linear unsteady, and nonlinear unsteady. 
The quasi-steady baseline keeps the noncirculatory term from Eq.~(\ref{eq:wagner_lift}) but replaces the circulatory convolution by the instantaneous steady value $2\pi w_{3/4}$. 
The linear Wagner baseline uses Eqs.~(\ref{eq:wagner_lift})--(\ref{eq:wagner_phi}). The nonlinear Wagner baseline replaces $w_{3/4}$ in the circulatory term by
\begin{equation}
    w_{3/4}^{\mathrm{eff}} = w_{3/4}\left[1 - 0.1\left(\frac{2\pi\,w_{3/4}}{U}\right)^{\!2}\right].
    \label{eq:nl_correction}
\end{equation}
All Wagner baselines are evaluated on the same $\tau$ grid and motion histories as the CFD data. The raw Wagner lift coefficient is denoted by $c_{l,\mathrm{W}}^{0}(t)$, which is calculated using incompressible assumptions and does not capture the steady lift offset caused by finite Mach number, airfoil shape, and nonzero mean angle of attack. Therefore, Eq.~(\ref{eq:wagner_mean_alignment}) defines a scalar static correction before the unsteady residual is calculated. For each motion case, this correction is
\begin{align}
    \Delta c_{l,\mathrm{static}}
    &=
    \mathrm{mean}\!\left(c_{l,\mathrm{CFD}}\right)
    -
    \mathrm{mean}\!\left(c_{l,\mathrm{W}}^{0}\right), \nonumber \\
    c_{l,\mathrm{W}}(t)
    &=
    c_{l,\mathrm{W}}^{0}(t)
    +
    \Delta c_{l,\mathrm{static}} .
    \label{eq:wagner_mean_alignment}
\end{align}
The term $c_{l,\mathrm{W}}(t)$ denotes the Wagner baseline after the static correction. This residual target is then the remaining time-dependent difference between the CFD lift and the lift obtained with the static-corrected Wagner baseline. In this study, the available CFD history is used to define the scalar offset in Eq.~(\ref{eq:wagner_mean_alignment}) for each benchmark case, and the static correction was also fitted from motion and flow parameters, including Mach number, mean and amplitude of the prescribed kinematics, reduced frequency, and the mean raw Wagner lift. This fit captured the available offsets sufficiently for this study, but in future applications, $\Delta c_{l,\mathrm{static}}$ could be supplied by a static ROM trained from static CFD runs, allowing the unsteady residual ROM to focus only on the unsteady correction.

\begin{figure}[htbp]
    \centering
    \subfigure[Pitch only]{%
        \begin{minipage}{0.48\linewidth}
            \centering
            \includegraphics[width=\textwidth]{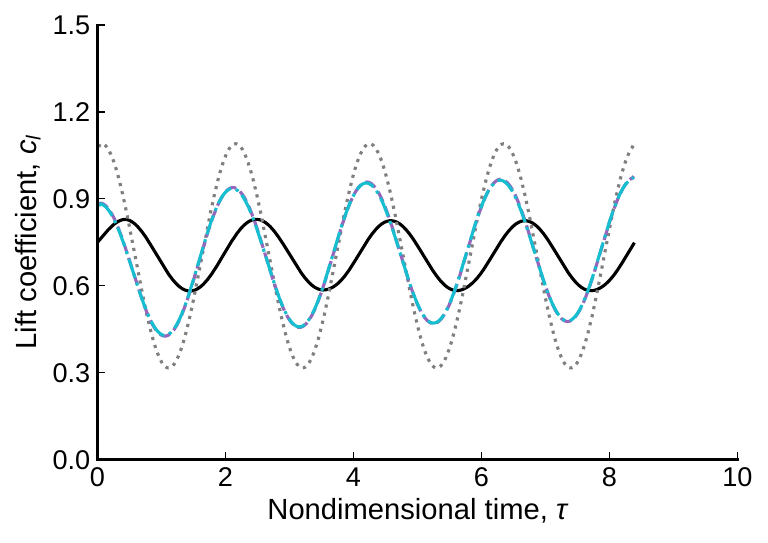}
        \end{minipage}%
    }\hspace{2mm}
    \subfigure[Plunge only]{%
        \begin{minipage}{0.48\linewidth}
            \centering
            \includegraphics[width=\textwidth]{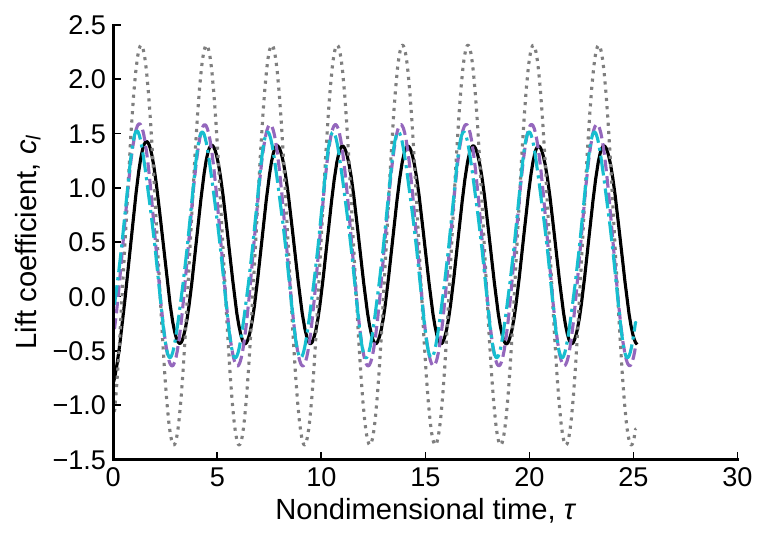}
        \end{minipage}%
    }\\
    \subfigure[Coupled pitch--plunge]{%
        \begin{minipage}{0.48\linewidth}
            \centering
            \includegraphics[width=\textwidth]{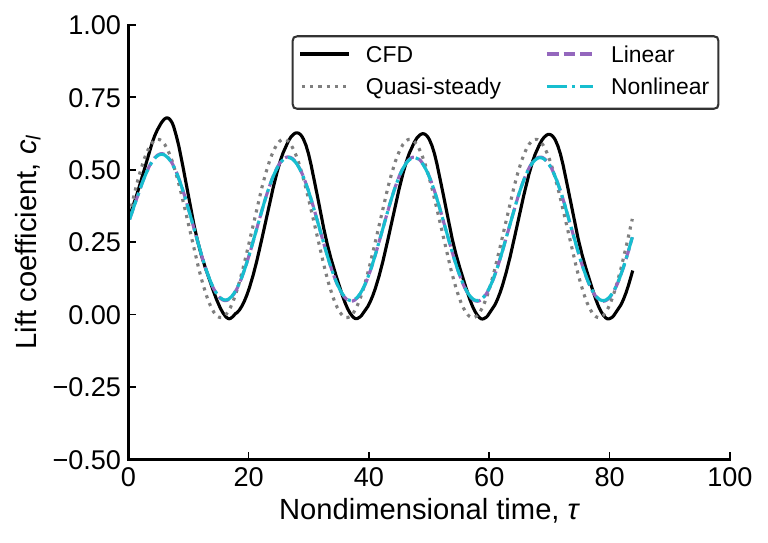}
        \end{minipage}%
    }
    \vspace{2mm}
    \caption{Quasi-steady, linear Wagner, and nonlinear Wagner comparison on external benchmark cases, corresponding to NRMSE values in Table~\ref{t:baseline_selection}.}
    \label{fr:wagner_comparison}
\end{figure}

Table~\ref{t:baseline_selection} and Fig.~\ref{fr:wagner_comparison} compare the candidate analytical baselines on the three external benchmark cases.
\begin{table}
    \centering
    \caption{Analytical baseline NRMSE for external benchmark cases.}
    \vspace{3mm}
    \begin{tabular}{lrrccc}
    \toprule
    Case & $M$ & $k$ & \shortstack{Quasi-\\steady} & \shortstack{Linear\\unsteady} & \shortstack{Nonlinear\\unsteady} \\
    \midrule
    Pitch only & 0.705 & 3.01 & 0.964 & 0.658 & 0.655 \\
    Plunge only & 0.720 & 2.00 & 0.377 & 0.240 & 0.260 \\
    Coupled pitch--plunge  & 0.750 & 0.30 & 0.160 & 0.114 & 0.116 \\
    \midrule
    Mean        &       &      & 0.500 & 0.337 & 0.344 \\
    \bottomrule
    \end{tabular}
    \label{t:baseline_selection}
\end{table}
Table~\ref{t:baseline_selection} reports the NRMSE values, while Fig.~\ref{fr:wagner_comparison} shows the corresponding CFD and baseline time histories. The quasi-steady baseline has the largest errors because it lacks the circulatory memory and phase lag contributions. The linear and nonlinear unsteady predictions are nearly identical for these cases, indicating that the motions remain within a regime where the linear unsteady response dominates and the nonlinear amplitude correction has no significant effect. Linear unsteady Wagner is therefore selected as the physics-based baseline for all subsequent residual learning studies.

\section{Residual Learning Methodology}
\label{s3}

This section defines the residual-learning formulation, describes the direct model and the residual model based on the analytical baseline (linear unsteady Wagner), and specifies the training, evaluation, and feature-selection procedures used for all generalization studies.

\subsection{Residual-Learning Model Formulation}
\label{s3-1}

In this study, a general residual-learning framework is proposed which decomposes the desired aerodynamic output into a physics-based baseline and one or more residual modules:
\begin{equation}
    \bm{F}_{\mathrm{total}}
    =
    \bm{F}_{\mathrm{baseline}}^{(f_0)}
    +
    \sum_{i=1}^{N} g_i \Delta \bm{F}_i^{(f_i)},
    \qquad
    0 \leq g_i \leq 1 .
    \label{eq:general_residual}
\end{equation}
Here, $\Delta \bm{F}_i^{(f_i)}$ is a learned residual correction associated with a chosen physical effect or fidelity level, and $g_i$ is an optional activation or blending factor. Although Eq.~(\ref{eq:general_residual}) defines a general residual-learning framework, the present work evaluates only the single-module unsteady lift case, with one LSTM correction, fixed $g_1=1$, and the linear Wagner model as the baseline for $c_l$. Broader multi-module applications will be explored in future work.

The notation below reports the model output at a generic time $t$, but the implemented sequences are given in the nondimensional coordinate $\tau$ defined in Eq.~(\ref{eq:xflow_time}). Each LSTM operates on the input sequence history available up to the current sample.
The direct model prediction is obtained from the input sequence as
\begin{equation}
    c_{l,\mathrm{direct}}(t)
    =
    \mathrm{LSTM}_{\mathrm{direct}}\left(\bm{x}_{\mathrm{direct}}(t)\right),
    \label{eq:direct_lstm}
\end{equation}
with the selected direct input set
\begin{equation}
    \bm{x}_{\mathrm{direct}}(t)
    =
    \begin{bmatrix}
    \alpha(t) & \dot{\alpha}(t) & \dot{h}(t) & M
    \end{bmatrix}^{T} .
    \label{eq:direct_final_input}
\end{equation}

The residual model learns the difference between the CFD lift and the Wagner baseline,
\begin{equation}
    \Delta c_l(t)
    =
    c_{l,\mathrm{CFD}}(t)-c_{l,\mathrm{W}}(t),
    \label{eq:residual_target}
\end{equation}
where this residual is the training target for the residual LSTM. The residual LSTM output is
\begin{equation}
    \Delta c_{l,\mathrm{pred}}(t)
    =
    \mathrm{LSTM}_{\mathrm{res}}\left(\bm{x}_{\mathrm{res}}(t)\right).
    \label{eq:res_lstm}
\end{equation}
The final residual-model lift prediction is obtained by adding the predicted residual correction to the Wagner baseline,
\begin{equation}
    c_{l,\mathrm{res}}(t)
    =
    c_{l,\mathrm{W}}(t)
    +
    \Delta c_{l,\mathrm{pred}}(t).
    \label{eq:res_reconstruction}
\end{equation}
The selected residual input set uses derivative and acceleration features along with Mach number,
\begin{equation}
    \bm{x}_{\mathrm{res}}(t)
    =
    \begin{bmatrix}
    \dot{h}(t) & \ddot{h}(t) & \alpha(t) & \dot{\alpha}(t) & \ddot{\alpha}(t) & M
    \end{bmatrix}^{T} .
    \label{eq:res_final_input}
\end{equation}
The feature and normalization checks used to select these direct and residual input sets are discussed in Sec.~\ref{s3-3}.

\subsection{Training Procedure and Error Metrics}
\label{s3-2}

Both direct and residual models use the same LSTM architecture. The network contains one LSTM layer with sequence output, a fully connected layer, and a regression output layer. The selected architecture and training settings are summarized in Table~\ref{t:selected_hyperparameters}.
\begin{table}
    \centering
    \caption{Selected LSTM architecture and training settings.}
    \vspace{3mm}
    \begin{tabular}{llp{0.42\textwidth}}
    \toprule
    Parameter & Value & Note \\
    \midrule
    Architecture & 1-layer LSTM & Consistent with the prior NLR 7301 LSTM setup~\cite{Wang2020}. \\
    Hidden size & 250 & Larger hidden size did not improve the priority cases. \\
    Optimizer & Adam & Learning rate $10^{-3}$. \\
    Epoch handling & 1000, restore-best & Selected using validation reconstructed-$c_l$ RMSE. \\
    Loss & MSE & Applied to the selected training target for each model. \\
    \bottomrule
    \end{tabular}
    \label{t:selected_hyperparameters}
\end{table}
All training, validation, and testing are performed on the original $\tau$-indexed solver samples, without conversion to physical time or interpolation to a common time step. All splits are made by complete motion case rather than by time sample, and the external benchmark cases are never used for training or validation. For residual models, reported errors are always computed after reconstructing $c_l$ using Eq.~\eqref{eq:res_reconstruction} and the best epoch is selected with the lowest validation reconstructed-$c_l$ RMSE. Since LSTM training can depend on random initialization, direct and residual models are trained with five independent random seeds for all the studies in this paper. 

Two error metrics are used to evaluate the reconstructed lift histories. Let $N_e$ be the number of samples in the evaluation window, and let $c_{l,\mathrm{pred}}$ denote either $c_{l,\mathrm{direct}}$ or $c_{l,\mathrm{res}}$. The root-mean-square error is
\begin{equation}
    \mathrm{RMSE}
    =
    \sqrt{
    \frac{1}{N_e}
    \sum_{n=1}^{N_e}
    \left(
    c_{l,\mathrm{CFD}}(t_n)-c_{l,\mathrm{pred}}(t_n)
    \right)^2
    },
    \label{eq:rmse}
\end{equation}
and the normalized RMSE is computed using the CFD lift range over the same samples,
\begin{equation}
    \mathrm{NRMSE}
    =
    \frac{\mathrm{RMSE}}
    {\max(c_{l,\mathrm{CFD}})-\min(c_{l,\mathrm{CFD}})} .
    \label{eq:nrmse}
\end{equation}
In Eq.~(\ref{eq:nrmse}), the maximum and minimum are taken over the same evaluation window used in Eq.~(\ref{eq:rmse}). Unless otherwise stated, the evaluation window is the final 60\% of each time history so that zero-state startup transients do not dominate the comparison.

\subsection{Feature Selection}
\label{s3-3}

The feature and normalization study identifies the input representation and target scaling used for the final direct and residual models. This check is needed because the direct model learns the full lift coefficient, while the residual model learns the remaining correction after subtracting the Wagner baseline. These two different targets can favor different kinematic variables and scaling choices. The entries in Tables~\ref{t:residual_ablation} and~\ref{t:direct_feature_ablation} are NRMSE values on the three external benchmark cases, and the final column is the mean across those cases. In Table~\ref{t:residual_ablation}, X-only denotes input normalization only, while X/Y denotes normalization of both the inputs and the residual target. Table~\ref{t:residual_ablation} shows that the best residual model uses the derivative and acceleration feature set $[\dot{h},\ddot{h},\alpha,\dot{\alpha},\ddot{\alpha},M]^T$ with input-only normalization. Adding Mach to the derivative and acceleration features improves the result across all cases, and not normalizing the residual target reduces the mean NRMSE from 0.060 to 0.044 for the same feature set. 
\begin{table}
    \centering
    \caption{Residual-model external-benchmark NRMSE for selected feature-set and target-normalization cases.}
    \vspace{3mm}
    \begin{tabular}{llrrrr}
    \toprule
    Feature set & Normalization & Pitch-only & Plunge-only & Coupled & Mean \\
    \midrule
    $\alpha$, $\dot\alpha$, $\dot h$, $M$
        & X-only & 0.124 & 0.035 & 0.091 & 0.083 \\
    $\dot h$, $\ddot h$, $\alpha$, $\dot\alpha$, $\ddot\alpha$
        & X-only & 0.063 & 0.043 & 0.089 & 0.065 \\
    $\dot h$, $\ddot h$, $\alpha$, $\dot\alpha$, $\ddot\alpha$, $M$
        & X/Y    & 0.078 & 0.028 & 0.075 & 0.060 \\
    \textbf{$\dot h$, $\ddot h$, $\alpha$, $\dot\alpha$, $\ddot\alpha$, $M$}
        & \textbf{X-only} & \textbf{0.058} & \textbf{0.036} & \textbf{0.039} & \textbf{0.044} \\
    $\alpha$, $h$, $\dot\alpha$, $\dot h$, $\ddot\alpha$, $\ddot h$, $M$
        & X-only & 0.066 & 0.030 & 0.128 & 0.075 \\
    \bottomrule
    \end{tabular}
    \label{t:residual_ablation}
\end{table}
Table~\ref{t:direct_feature_ablation} reports the corresponding feature study for the direct model. The direct model performs best with the compact rate-based input set $[\alpha,\dot{\alpha},\dot{h},M]^T$ and a normalized output. Adding second-derivative features raises the mean NRMSE from 0.054 to 0.104. This indicates that the direct and residual models benefit from different representations. The residual model benefits from variables aligned with the Wagner function, while the direct model is more sensitive to noisy acceleration inputs. Similarly, the selected direct model benefits from using normalization for total $c_l$, whereas the residual model benefits by normalizing the input only since it preserves the physical scale of $c_{l,\mathrm{CFD}}-c_{l,\mathrm{W}}$.

\begin{table}
    \centering
    \caption{Direct-model external-benchmark NRMSE for selected feature sets.}
    \vspace{3mm}
    \begin{tabular}{lrrrr}
    \toprule
    Feature set & Pitch-only & Plunge-only & Coupled & Mean \\
    \midrule
    $\alpha$, $h$, $\dot\alpha$, $\dot h$                               & 0.070 & 0.042 & 0.122 & 0.078 \\
    \textbf{$\alpha$, $\dot\alpha$, $\dot h$, $M$}                      & \textbf{0.041} & \textbf{0.049} & \textbf{0.072} & \textbf{0.054} \\
    $\alpha$, $h$, $\dot\alpha$, $\dot h$, $M$                          & 0.063 & 0.021 & 0.106 & 0.063 \\
    $\dot h$, $\ddot h$, $\alpha$, $\dot\alpha$, $\ddot\alpha$, $M$     & 0.123 & 0.026 & 0.164 & 0.104 \\
    $\alpha$, $h$, $\dot\alpha$, $\dot h$, $\ddot\alpha$, $\ddot h$, $M$ & 0.050 & 0.021 & 0.105 & 0.059 \\
    \bottomrule
    \end{tabular}
    \label{t:direct_feature_ablation}
\end{table}

\section{Results}
\label{s4}

This section presents the external benchmark and generalization results. It first evaluates the direct and residual models on the external benchmark cases and then examines leave-one-out and leave-family-out generalization across the four motion families.

\subsection{External Benchmark Performance}
\label{s4-1}

The external benchmark study evaluates the selected direct and residual models on three test cases that are not used during training or validation. A random-seed sensitivity analysis is used to check whether the comparison is stable across different random initializations. Table~\ref{t:external_seed_summary} reports the mean and standard deviation of the NRMSE over five random seeds for each model and case. The residual model gives a lower mean error over the three cases, with $0.054 \pm 0.007$ compared with $0.065 \pm 0.009$ for the direct model. However, it is also important to note that the case-wise behavior is more informative than the mean values. The direct model remains better for the pitch-only case, while the residual model is better for the plunge-only and coupled pitch-plunge cases.
\begin{table}
    \centering
    \caption{External-test NRMSE over five random seeds.}
    \vspace{3mm}
    \begin{tabular}{lrrrr}
    \toprule
    Model & Pitch-only & Plunge-only & Coupled & Mean \\
    \midrule
    Direct & $0.052 \pm 0.020$ & $0.056 \pm 0.011$ & $0.086 \pm 0.024$ & $0.065 \pm 0.009$ \\
    Residual & $0.070 \pm 0.016$ & $0.026 \pm 0.010$ & $0.066 \pm 0.010$ & $0.054 \pm 0.007$ \\
    \bottomrule
    \end{tabular}
    \label{t:external_seed_summary}
\end{table}

\begin{figure}
    \centering
    \includegraphics[width=0.68\textwidth]{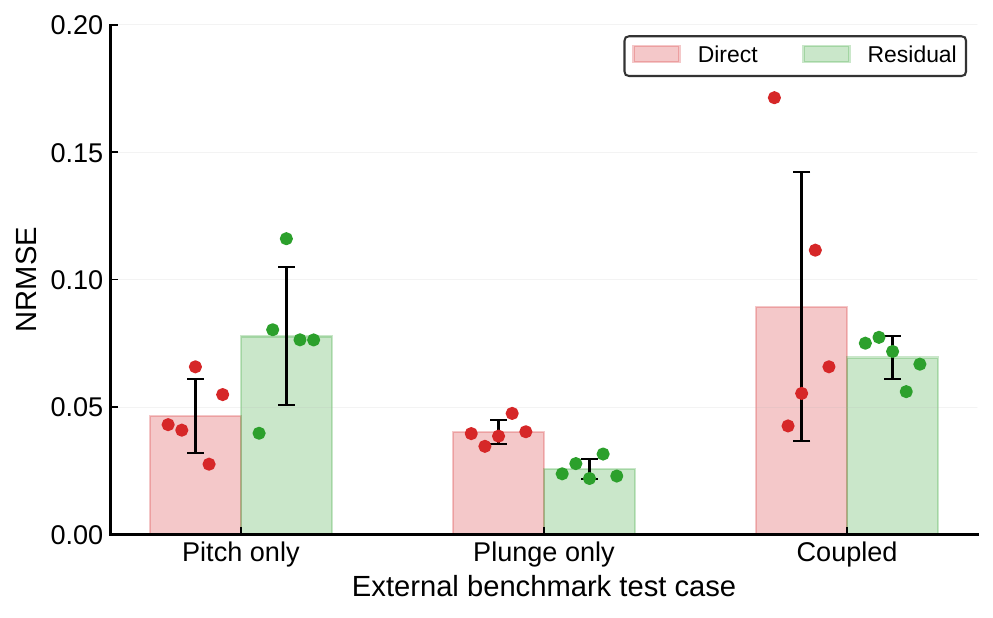}
    \vspace{2mm}
    \caption{External-test NRMSE comparison across random seeds.}
    \label{fr:seed_robustness}
\end{figure}

Figure~\ref{fr:seed_robustness} shows the same external benchmark errors across the five random seeds. The bars show means over the five random seeds and the dots show individual random-seed results. The residual model has smaller scatter across random seeds on the plunge-only and coupled cases. It also shows that the pitch-only case is where the direct model remains preferable because, for this external pure-pitch case, the Wagner baseline does not reduce the phase and amplitude mismatch enough to make the residual target easier than direct lift prediction.

\begin{figure}
    \centering
    \subfigure[Lift coefficient]{%
        \begin{minipage}{0.48\linewidth}
            \centering
            \includegraphics[width=\textwidth]{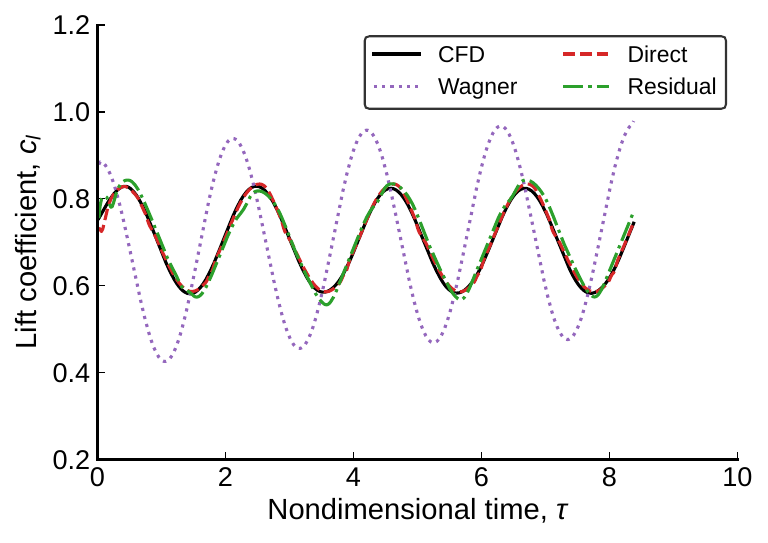}
        \end{minipage}%
    }\hspace{2mm}
    \subfigure[Residual correction]{%
        \begin{minipage}{0.48\linewidth}
            \centering
            \includegraphics[width=\textwidth]{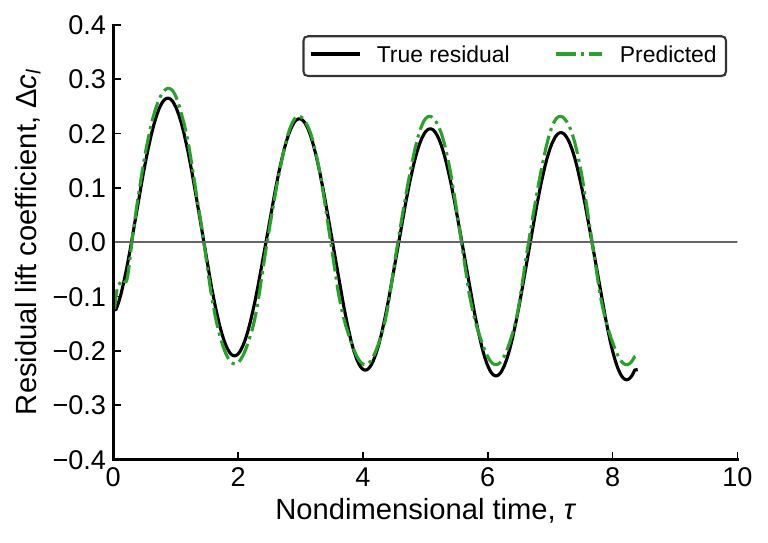}
        \end{minipage}%
    }\\
    \subfigure[Prediction error]{%
        \begin{minipage}{0.48\linewidth}
            \centering
            \includegraphics[width=\textwidth]{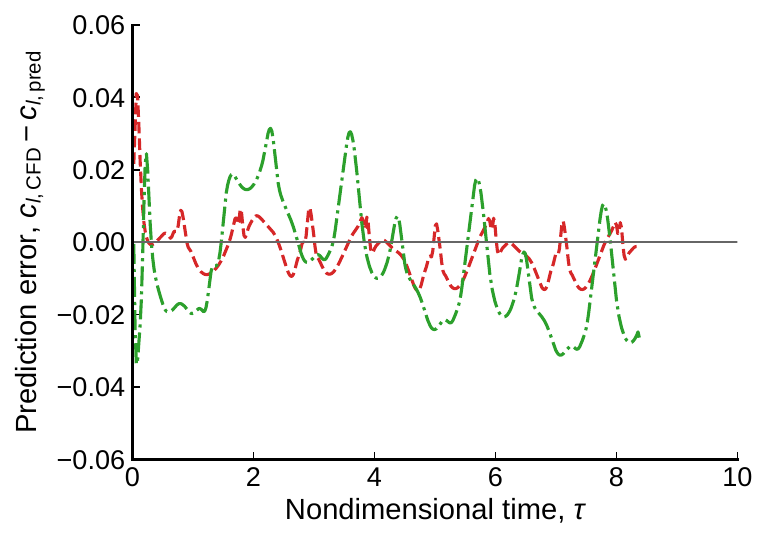}
        \end{minipage}%
    }
    \vspace{2mm}
    \caption{Pitch-only external benchmark test case comparison.}
    \label{fr:median_time_history_case02}
\end{figure}

\begin{figure}
    \centering
    \subfigure[Lift coefficient]{%
        \begin{minipage}{0.48\linewidth}
            \centering
            \includegraphics[width=\textwidth]{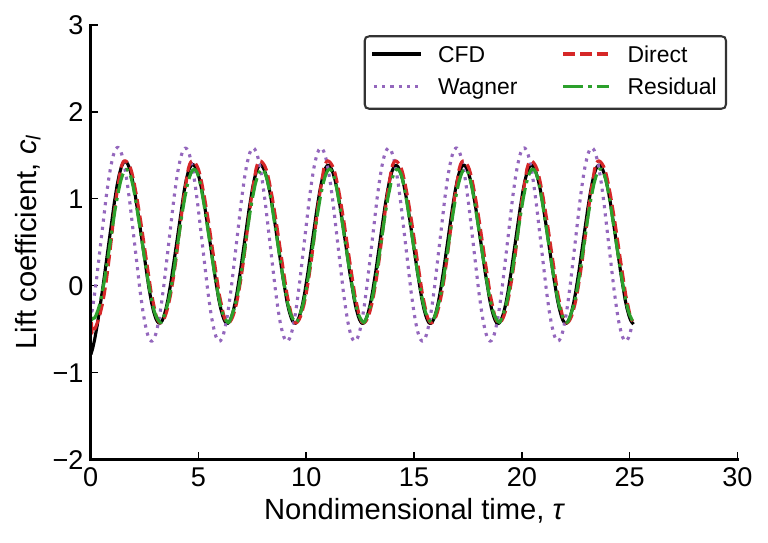}
        \end{minipage}%
    }\hspace{2mm}
    \subfigure[Residual correction]{%
        \begin{minipage}{0.48\linewidth}
            \centering
            \includegraphics[width=\textwidth]{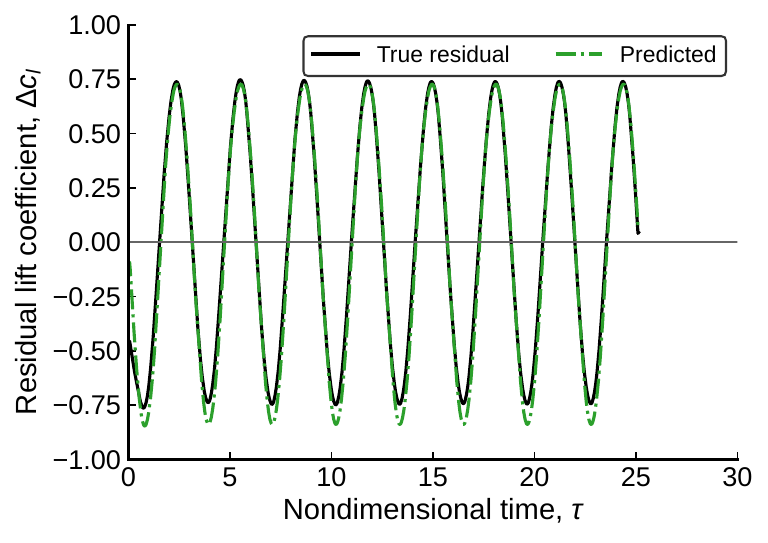}
        \end{minipage}%
    }\\
    \subfigure[Prediction error]{%
        \begin{minipage}{0.48\linewidth}
            \centering
            \includegraphics[width=\textwidth]{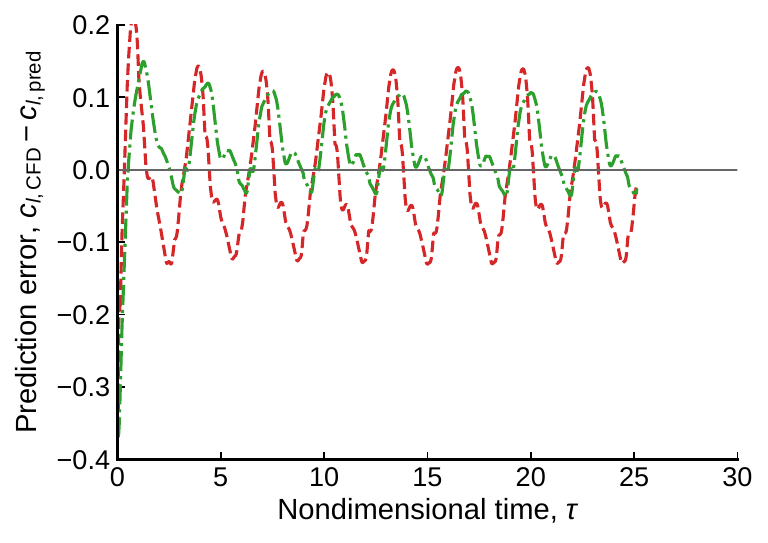}
        \end{minipage}%
    }
    \vspace{2mm}
    \caption{Plunge-only external benchmark test case comparison.}
    \label{fr:median_time_history_case10}
\end{figure}

\begin{figure}
    \centering
    \subfigure[Lift coefficient]{%
        \begin{minipage}{0.48\linewidth}
            \centering
            \includegraphics[width=\textwidth]{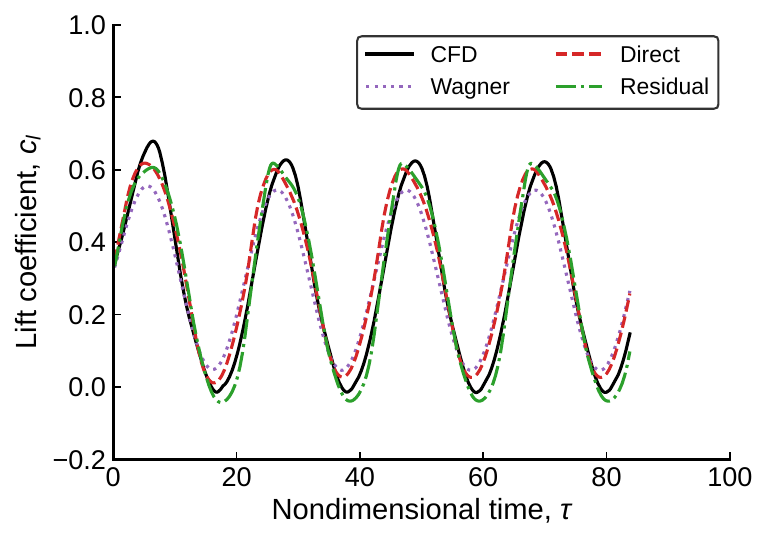}
        \end{minipage}%
    }\hspace{2mm}
    \subfigure[Residual correction]{%
        \begin{minipage}{0.48\linewidth}
            \centering
            \includegraphics[width=\textwidth]{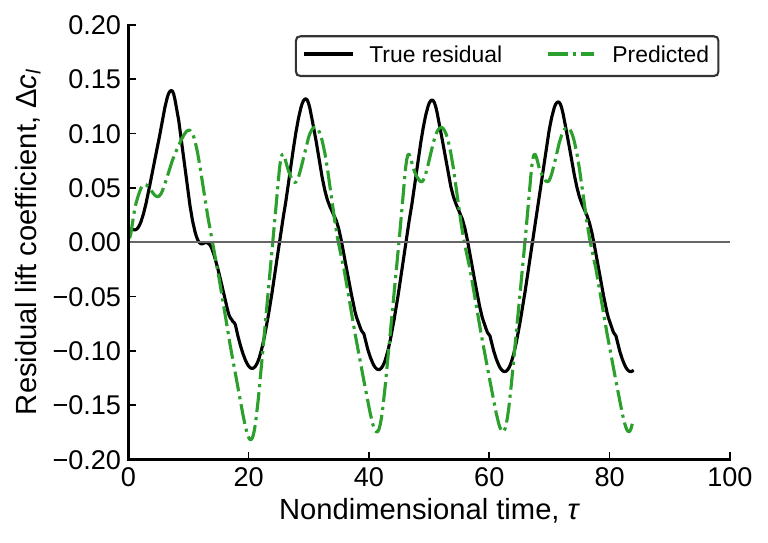}
        \end{minipage}%
    }\\
    \subfigure[Prediction error]{%
        \begin{minipage}{0.48\linewidth}
            \centering
            \includegraphics[width=\textwidth]{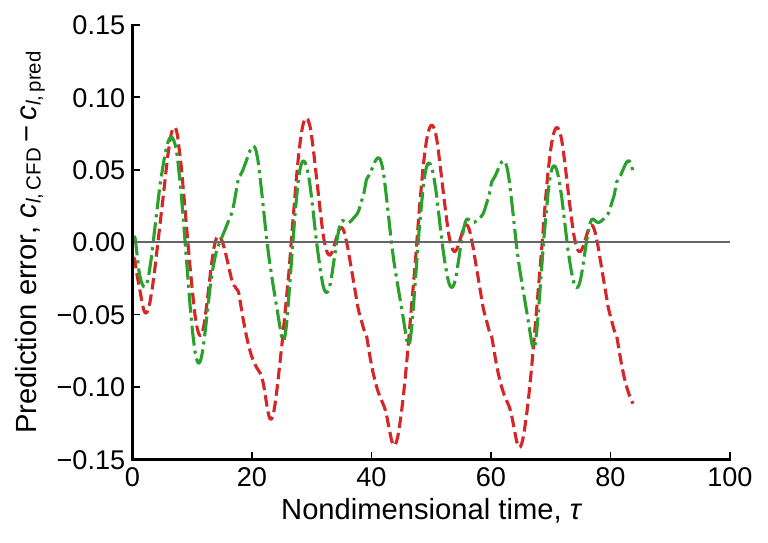}
        \end{minipage}%
    }
    \vspace{2mm}
    \caption{Coupled pitch--plunge external benchmark test case comparison.}
    \label{fr:median_time_history_case15}
\end{figure}

The time-history comparisons provide more insight into the different performance of the direct and residual models across the three cases. The time histories are shown for the median seed in each case, but similar trends are observed across all seeds. Figure~\ref{fr:median_time_history_case02} compares the reconstructed lift, residual correction, and pointwise prediction error for the pitch-only external benchmark case. The time-history comparison shows that both models follow the lift oscillation, but the direct model gives a smaller NRMSE since the residual model needs to learn both the phase and amplitude difference between the baseline and the CFD. 
Figure~\ref{fr:median_time_history_case10} shows the plunge-only external benchmark case. The reconstructed lift histories show closer agreement for the residual model. This is the regime where the Wagner baseline removes a useful part of the memory and phase behavior, leaving the LSTM to learn a smaller correction.
Next, Fig.~\ref{fr:median_time_history_case15} shows the coupled pitch--plunge external benchmark case. The residual prediction follows the CFD lift more closely than the direct model, and the pointwise error is smaller over most of the evaluation window. This result suggests that the Wagner baseline provides useful phase and memory structure for the low-frequency coupled motion, even though the baseline alone does not fully match the CFD response. Overall, both models are less accurate than in the plunge-only case. This is likely because the coupled case has a longer duration and a coarser nondimensional time-step, making the LSTM evaluation less accurate for this case.

\subsection{Leave-One-Out Generalization}
\label{s4-2}

The leave-one-out study removes one case at a time from a motion family and trains on the remaining cases. Test cases from the same family remain in the training set, so this study measures in-family interpolation to an unseen motion input. A random-seed sensitivity analysis was also conducted for this generalization study. Since the trends were similar across seeds, the results in this section are shown for one representative seed case.

\begin{figure}
    \centering
    \includegraphics[width=0.90\textwidth]{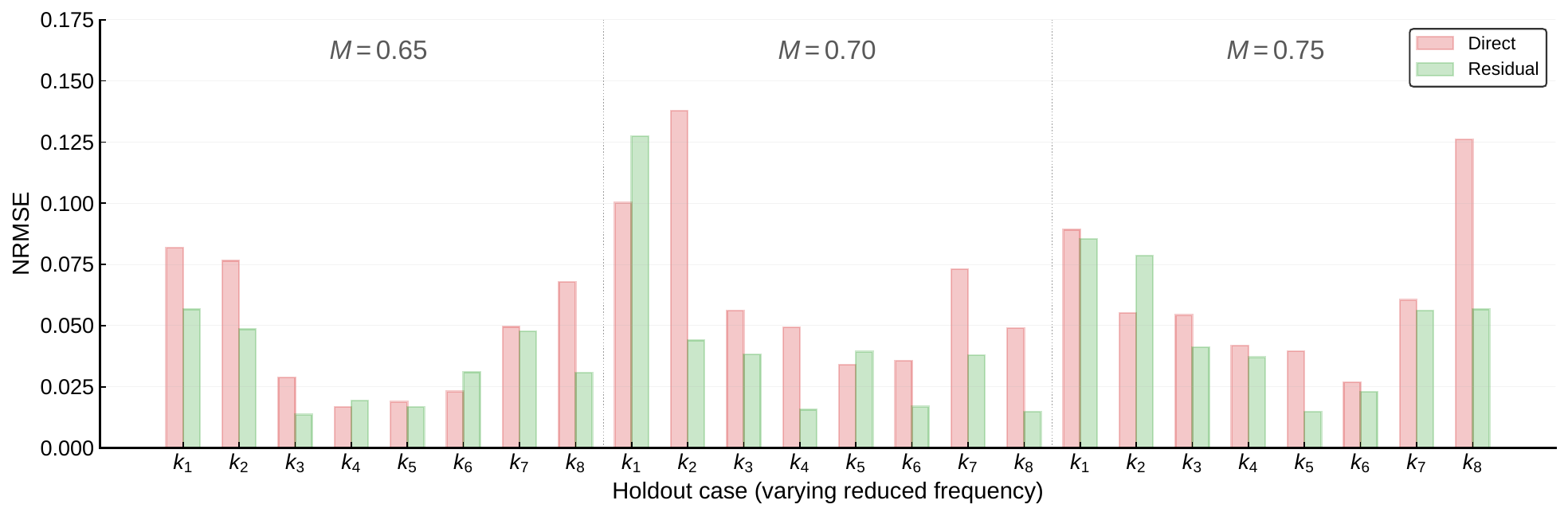}
    \vspace{2mm}
    \caption{Single-harmonic leave-one-out NRMSE by holdout case.}
    \label{fr:loo_sh}
\end{figure}

\begin{figure}
    \centering
    \includegraphics[width=0.49\textwidth]{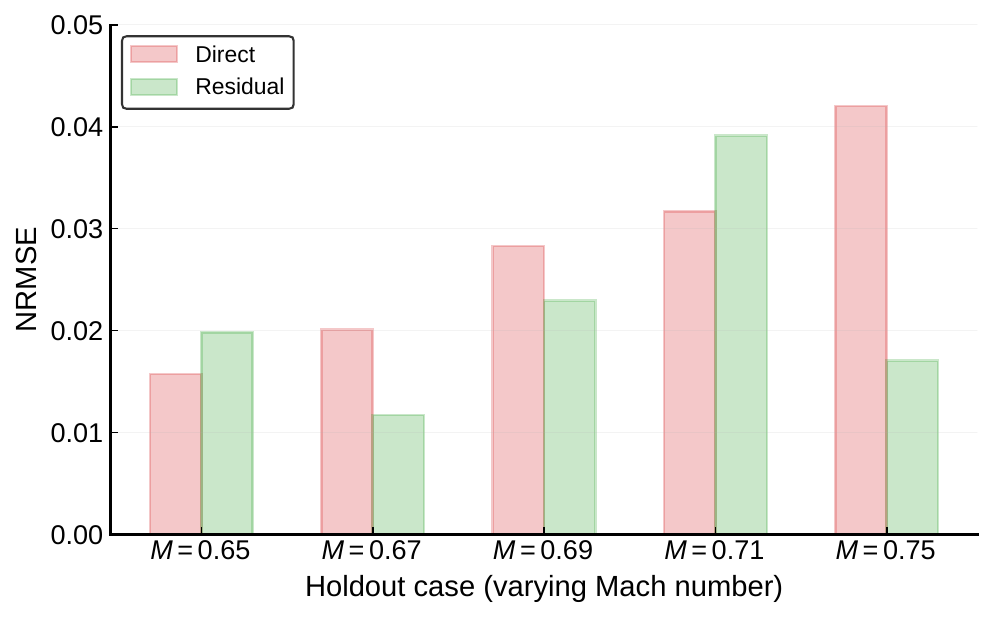}
    \vspace{2mm}
    \caption{Amplitude-modulated leave-one-out NRMSE by holdout case.}
    \label{fr:loo_am}
\end{figure}

\begin{figure}
    \centering
    \includegraphics[width=0.60\textwidth]{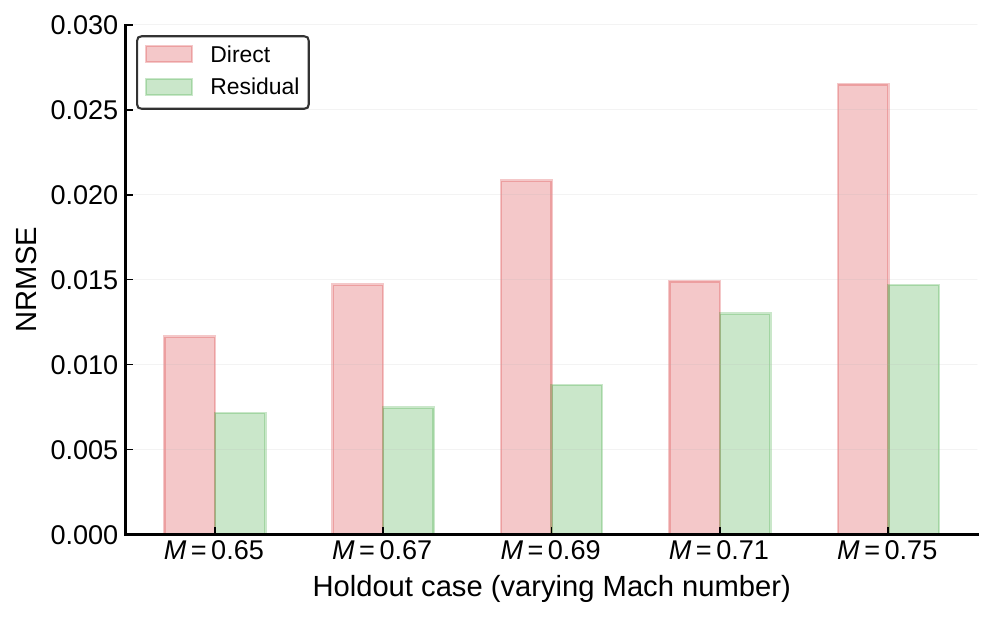}
    \vspace{2mm}
    \caption{Gaussian-pulse leave-one-out NRMSE by holdout case.}
    \label{fr:loo_gp}
\end{figure}

\begin{figure}
    \centering
    \includegraphics[width=0.96\textwidth]{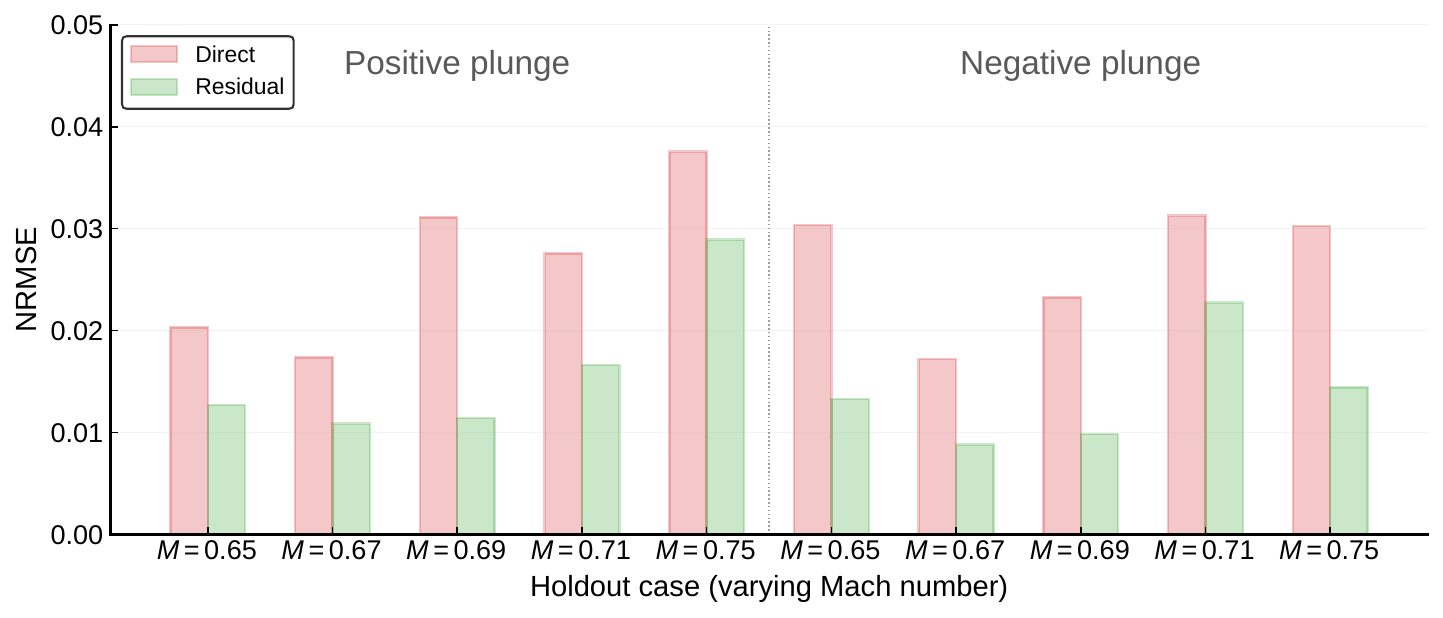}
    \vspace{2mm}
    \caption{Noise-perturbed leave-one-out NRMSE by holdout case.}
    \label{fr:loo_np}
\end{figure}

Figure~\ref{fr:loo_sh} shows the single-harmonic leave-one-out NRMSE by holdout case. These results help identify how the direct and residual models behave as reduced frequency and Mach number vary across the single-harmonic family. The residual model gives lower NRMSE for most of the single-harmonic holdouts. The largest gains occur mostly at higher reduced frequency and higher Mach number within the single-harmonic leave-one-out setting, where nearby training cases allow the LSTM to refine the phase structure supplied by the Wagner baseline. The direct model is more competitive for lower-frequency cases and outperforms the residual model. 
Next, Fig.~\ref{fr:loo_am} shows the amplitude-modulated harmonic leave-one-out results where the direct and residual errors are compared for each amplitude-modulated holdout case. The residual model has the lower family mean, although the direct model remains competitive at the lowest Mach number. This indicates that the residual formulation improves prediction for nonstationary harmonic inputs as well, but sparse Mach coverage within this family impacts the individual fold results.
Gaussian-pulse leave-one-out results are shown in Fig.~\ref{fr:loo_gp}; these results evaluate whether the residual advantage remains for transient inputs as well. The residual model gives lower errors for all Gaussian-pulse holdouts in this representative run. This suggests that the Wagner baseline still captures a useful part of the attached-flow response and leaves the LSTM a smaller correction to learn, even when the input is a short transient pulse rather than a sustained oscillation.
Next, Fig.~\ref{fr:loo_np} shows the noise-perturbed harmonic leave-one-out results. The figure separates the two plunge-sign variants and compares the Mach-number trend within each variant. The residual model gives lower NRMSE for every holdout in this family. This result indicates that the residual learning is not limited to unperturbed sinusoidal inputs and remains effective when the input contains broadband perturbations.

The results for the leave-one-out are summarized in Table~\ref{t:loo_summary} by motion family. The table reports the family-mean NRMSE and the number of holdouts for which the residual model is better than the direct model. The residual model has a lower mean for all four families. The amplitude-modulated family has fewer residual-favorable holdouts, but its residual-model mean remains lower. The table therefore supports the conclusion that the residual advantage extends beyond the sinusoidal training family.
\begin{table}
    \centering
    \caption{Leave-one-out generalization summary by motion family.}
    \vspace{3mm}
    \begingroup
    \setlength{\tabcolsep}{4pt}
    \renewcommand{\arraystretch}{1.12}
    \begin{tabularx}{\textwidth}{@{}>{\raggedright\arraybackslash}Xccc@{}}
    \toprule
    Motion family & Direct mean & Residual mean & Residual lower \\
    \midrule
    Single harmonic              & 0.057 & 0.045 & 19/24 \\
    Amplitude-modulated harmonic & 0.028 & 0.022 &  3/5  \\
    Gaussian pulse               & 0.018 & 0.010 &  5/5  \\
    Noise-perturbed harmonic     & 0.027 & 0.015 & 10/10 \\
    \bottomrule
    \end{tabularx}
    \endgroup
    \label{t:loo_summary}
\end{table}
Overall, across all four motion families, the residual model lowered the family-mean NRMSE, confirming that the linear unsteady Wagner baseline supplied a useful attached-flow response and left the LSTM a smaller correction to learn for in-family interpolation. The direct model remained competitive only at the lowest reduced frequencies and Mach numbers, where the Wagner baseline offered less benefit. 
\subsection{Leave-Family-Out Generalization}
\label{s4-3}

The final generalization study is the leave-family-out study, where an entire motion family is removed from both training and validation. This is a more challenging test than leave-one-out because the trained model has not seen any examples from the withheld family. The family holdout cases are used to evaluate generalization across motion type, while the external benchmark cases are evaluated with the same trained model. A random-seed sensitivity analysis was also conducted for this study. Similar to Sec.~\ref{s4-2}, the plots in this section are shown for one representative random-seed case. Since the Gaussian-pulse trend was not consistent across random-seed runs, Table~\ref{t:study8_summary} also compares the best and worst runs, defined by the residual improvement in family-mean NRMSE.

\begin{figure}
    \centering
    \subfigure[Family holdout cases]{%
        \begin{minipage}{0.96\linewidth}
            \centering
            \includegraphics[width=\textwidth]{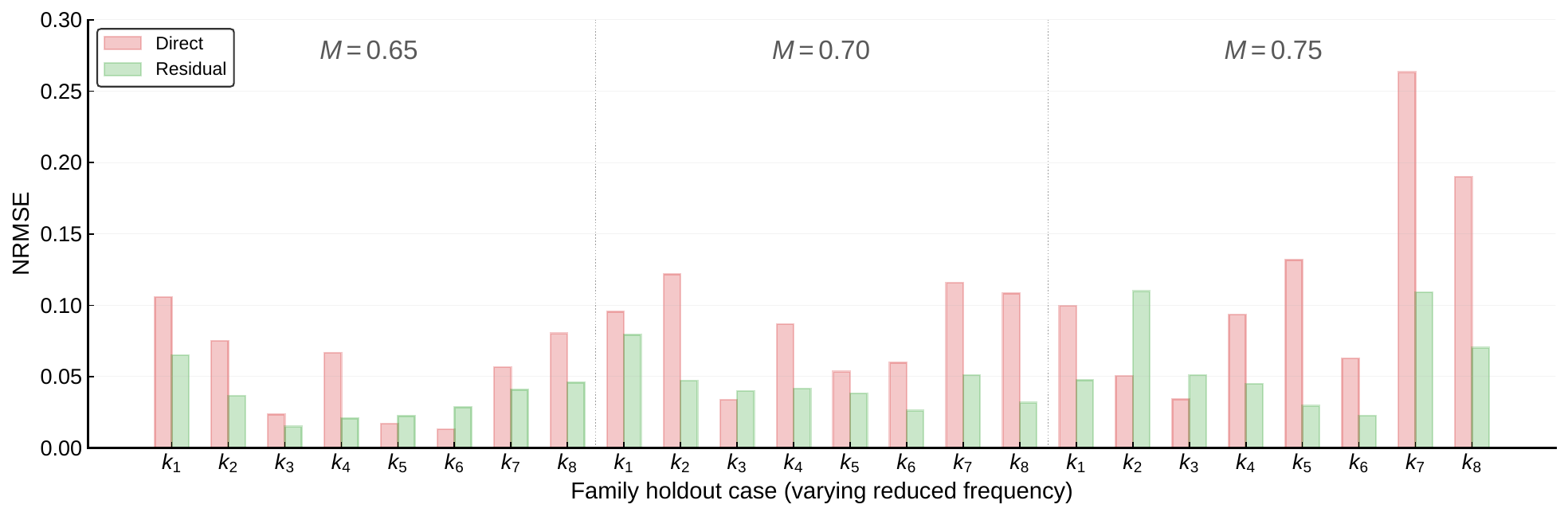}
        \end{minipage}%
    }\\
    \subfigure[External test cases]{%
        \begin{minipage}{0.60\linewidth}
            \centering
            \includegraphics[width=\textwidth]{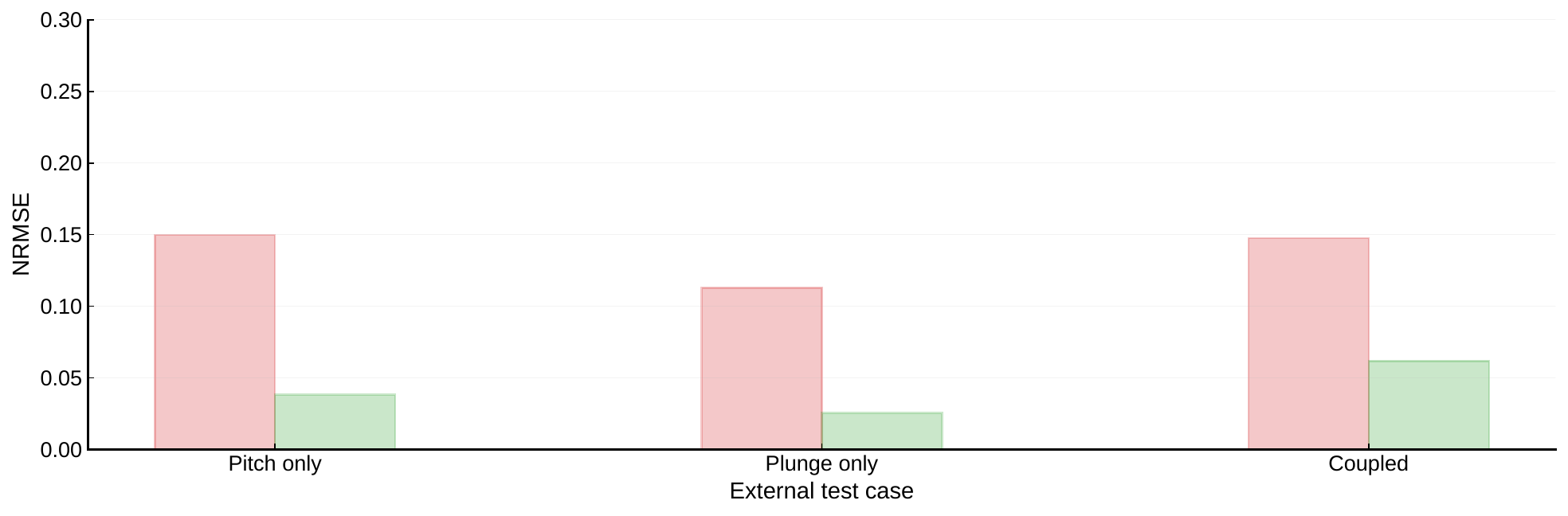}
        \end{minipage}%
    }
    \vspace{2mm}
    \caption{Single-harmonic leave-family-out NRMSE.}
    \label{fr:lfo_sh}
\end{figure}

\begin{figure}
    \centering
    \subfigure[Family holdout cases]{%
        \begin{minipage}{0.6\linewidth}
            \centering
            \includegraphics[width=\textwidth]{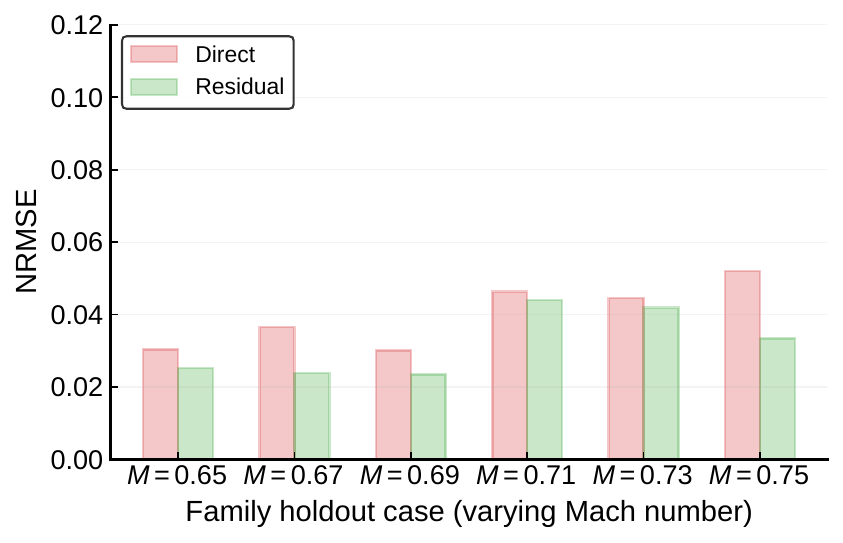}
        \end{minipage}%
    }\\
    \subfigure[External test cases]{%
        \begin{minipage}{0.60\linewidth}
            \centering
            \includegraphics[width=\textwidth]{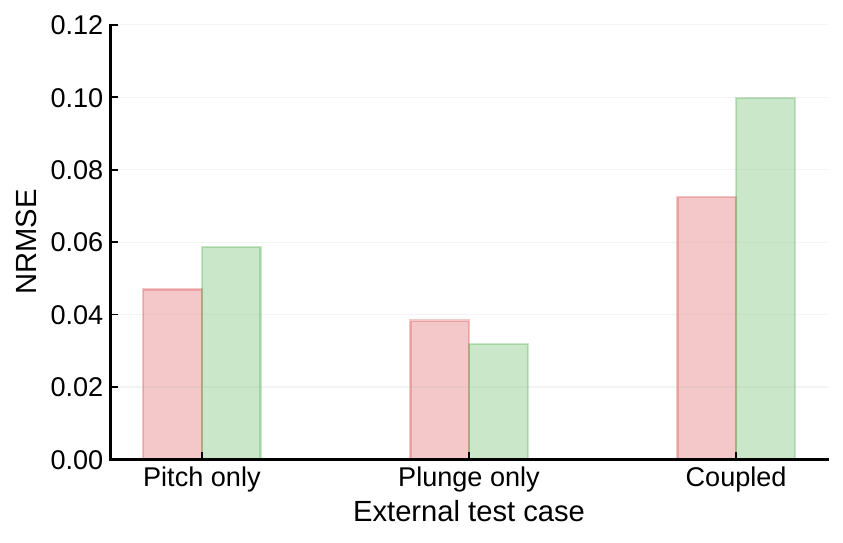}
        \end{minipage}%
    }
    \vspace{2mm}
    \caption{Amplitude-modulated leave-family-out NRMSE.}
    \label{fr:lfo_am}
\end{figure}

\begin{figure}
    \centering
    \subfigure[Family holdout cases]{%
        \begin{minipage}{0.96\linewidth}
            \centering
            \includegraphics[width=\textwidth]{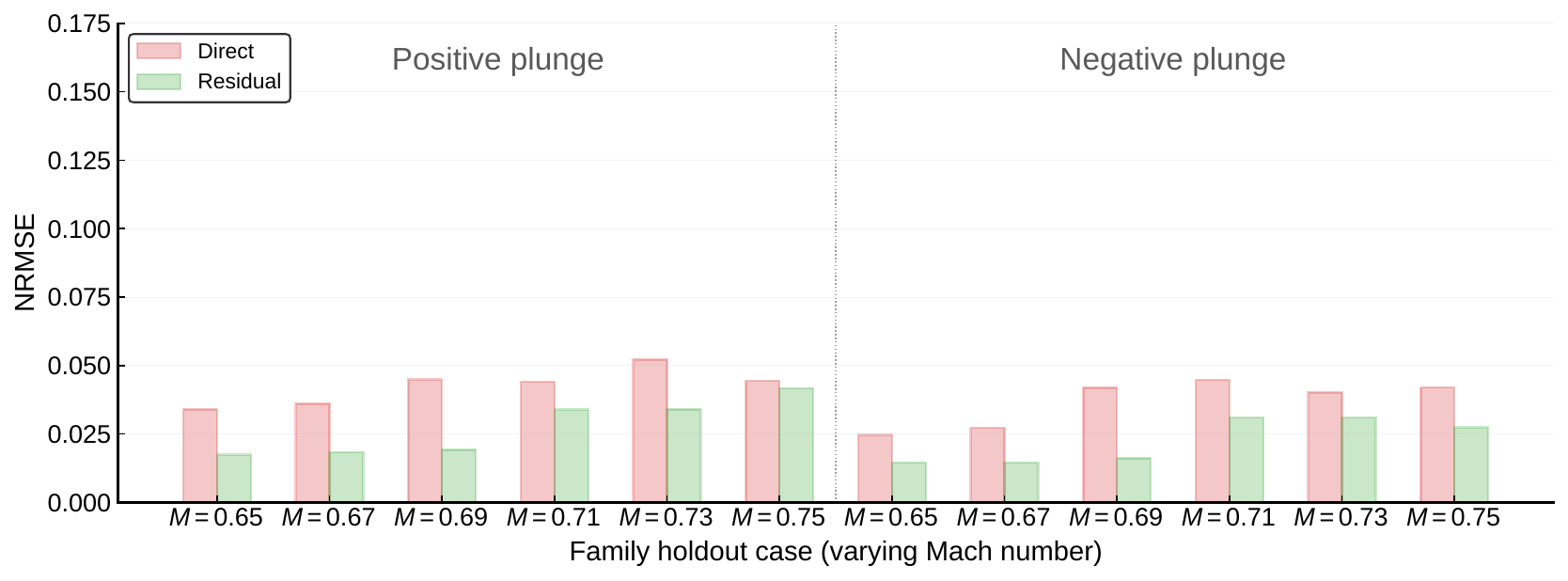}
        \end{minipage}%
    }\\
    \subfigure[External test cases]{%
        \begin{minipage}{0.60\linewidth}
            \centering
            \includegraphics[width=\textwidth]{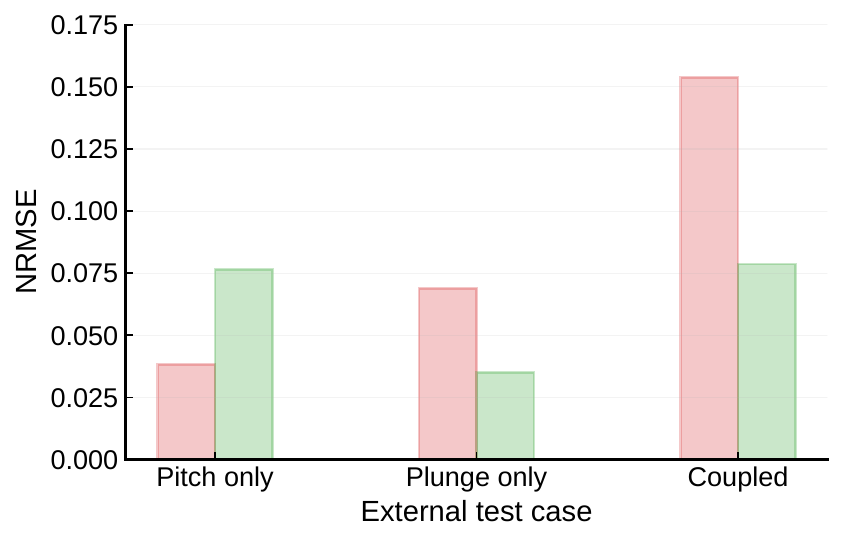}
        \end{minipage}%
    }
    \vspace{2mm}
    \caption{Noise-perturbed leave-family-out NRMSE.}
    \label{fr:lfo_np}
\end{figure}

Figure~\ref{fr:lfo_sh} shows the single-harmonic leave-family-out result. These results show that removing all single-harmonic cases from training mainly affects the direct model. The residual-model mean remains close to its leave-one-out value, indicating that the Wagner baseline supplies a transferable sinusoidal response structure while the direct model must learn the sinusoidal behavior only from non-sinusoidal examples.
Next, Fig.~\ref{fr:lfo_am} shows the amplitude-modulated harmonic leave-family-out result. For this motion family as well, the residual model has a lower error for every holdout case. This indicates that the residual formulation remains useful for slowly varying harmonic amplitudes, even when the training set contains no amplitude-modulated examples.
Next, Fig.~\ref{fr:lfo_np} shows the noise-perturbed harmonic leave-family-out result separated by the positive- and negative-plunge testing cases. The residual model gives lower NRMSE for all noise-perturbed holdouts as well. This indicates that residual learning is not limited to clean sinusoidal inputs and remains effective when the withheld family contains broadband perturbations.

\begin{figure}
    \centering
    \subfigure[Family holdout cases]{%
         \begin{minipage}{0.6\linewidth}
            \centering
            \includegraphics[width=\textwidth]{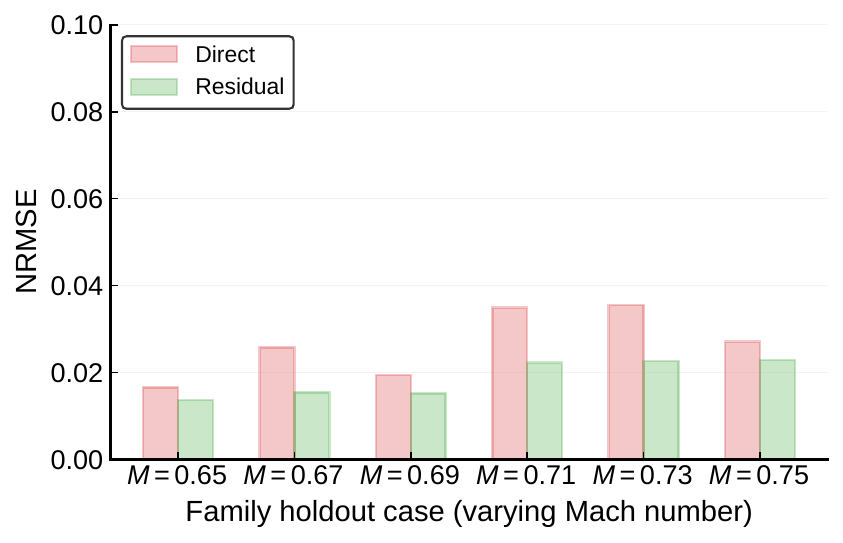}
        \end{minipage}%
    }\\
    \subfigure[External test cases]{%
        \begin{minipage}{0.60\linewidth}
            \centering
            \includegraphics[width=\textwidth]{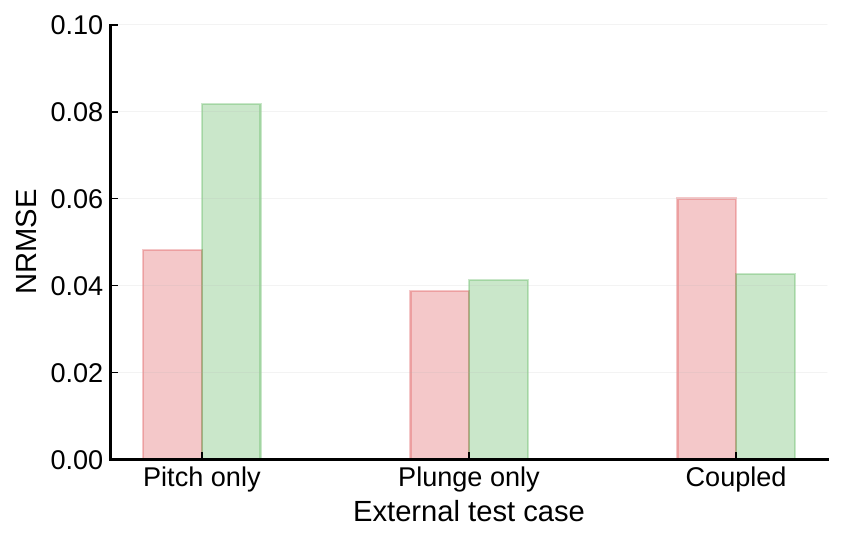}
        \end{minipage}%
    }
    \vspace{2mm}
    \caption{Gaussian-pulse leave-family-out NRMSE.}
    \label{fr:lfo_gp}
\end{figure}

Finally, Fig.~\ref{fr:lfo_gp} shows the Gaussian-pulse leave-family-out result. In the representative run shown here, the residual model has lower NRMSE for the Gaussian-pulse holdouts. However, the random-seed sensitivity analysis comparison in Table~\ref{t:study8_summary} shows that this is the most sensitive leave-family-out case, and the model behavior can change across random-seed runs.
Table~\ref{t:study8_summary} summarizes the leave-family-out results using the best and worst random-seed runs for each motion family. Best and worst are based on the reduction in family-mean NRMSE from the direct model to the residual model. The final column reports the number of family holdout cases, out of the family total, for which the residual model has the lower NRMSE. The table shows that the residual model remains better in family-mean NRMSE for the single-harmonic, amplitude-modulated, and noise-perturbed families even in the worst runs. The Gaussian-pulse family is the exception. The best run gives lower residual-model error for all holdouts, while the worst run favors the direct model for all holdouts. This indicates that the residual advantage is strongest when the Wagner baseline removes a transferable part of the response, but it can weaken for transient cases where the withheld motion differs more strongly from the remaining training families. 
\begin{table}
    \centering
    \caption{Leave-family-out generalization summary for random-seed runs.}
    \vspace{3mm}
    \begingroup
    \setlength{\tabcolsep}{4pt}
    \renewcommand{\arraystretch}{1.12}
    \begin{tabular}{@{}>{\raggedright\arraybackslash}p{0.34\textwidth}crrr@{}}
    \toprule
    Motion family & Run & Direct mean & Residual mean & Residual lower \\
    \midrule
    \multirow{2}{0.34\textwidth}{Single harmonic}
        & Best  & 0.091 & 0.046 & 19/24 \\
        & Worst & 0.085 & 0.046 & 19/24 \\
    \addlinespace
    \multirow{2}{0.34\textwidth}{Amplitude-modulated harmonic}
        & Best  & 0.045 & 0.032 & 6/6 \\
        & Worst & 0.040 & 0.032 & 6/6 \\
    \addlinespace
    \multirow{2}{0.34\textwidth}{Gaussian pulse}
        & Best  & 0.027 & 0.019 & 6/6 \\
        & Worst & 0.019 & 0.024 & 0/6 \\
    \addlinespace
    \multirow{2}{0.34\textwidth}{Noise-perturbed harmonic}
        & Best  & 0.042 & 0.022 & 12/12 \\
        & Worst & 0.040 & 0.025 & 12/12 \\
    \bottomrule
    \end{tabular}
    \endgroup
    \label{t:study8_summary}
\end{table}

\section{Concluding Remarks}
\label{s5}

This paper evaluated a residual learning framework for unsteady lift prediction on the NLR 7301 airfoil in the transonic flow regime in the presence of
shock motion. The residual framework uses the Wagner indicial function as an analytical baseline function. The study compared a direct LSTM trained on total CFD lift with a residual LSTM trained on the difference between the CFD lift and a mean-aligned Wagner prediction. The comparison used external benchmark tests, leave-one-out studies, and leave-family-out studies across single-harmonic, amplitude-modulated harmonic, Gaussian-pulse, and noise-perturbed harmonic motion families.

The baseline comparison and feature-selection study show that the direct and residual models favor different inputs and scaling choices. The linear Wagner model was selected because it performed nearly the same as the nonlinear Wagner variant and better than the quasi-steady baseline for the external benchmark cases. The residual model performed best with derivative and acceleration features and input-only normalization. The direct model performed best with a compact rate-based feature set and a normalized total-lift target. This difference indicates that the residual model is not simply a direct model with a redefined target. It is a lower-variance correction problem with its own useful feature representation. With these selected direct and residual models, the external benchmark results show that the residual model gives a lower mean NRMSE over five random seeds and smaller average scatter across random seeds than the direct model. However, the improvement is case dependent since the residual model performed better for the plunge-only and coupled pitch-plunge cases, where the Wagner gives a useful amplitude and phase baseline. The direct model remains better for the high-frequency pitch-only case, where the Wagner baseline does not simplify the residual target enough to improve the final reconstructed lift prediction.

The leave-one-out results show that the residual advantage is not limited to sinusoidal inputs. The residual model improves the family-mean NRMSE for all four motion families and gives lower NRMSE for most holdouts in the single-harmonic, Gaussian-pulse, and noise-perturbed families. The leave-family-out results provide a stronger generalization test because an entire motion family is removed from training. In these tests, the residual model is nearly unchanged when all single-harmonic cases are removed, while the direct model degrades substantially. The residual model also generalizes well to the amplitude-modulated and noise-perturbed family holdouts. The Gaussian-pulse family is the exception and the most sensitive case, suggesting that the residual advantage depends on how well the remaining training families represent the withheld transient response.

Overall, the results support residual learning as a useful way to augment classical unsteady aerodynamic models with data-driven corrections. The approach is most effective when the physics baseline removes a structured part of the aerodynamic response and leaves the neural network to learn a smaller unsteady correction. Future work should examine Mach-number interpolation and extrapolation more systematically, replace the present scalar static correction with a static ROM, and evaluate alternative baselines such as compressibility-corrected Wagner and other low-fidelity unsteady solvers. The framework should also be evaluated for other aerodynamic coefficients and additional test cases, including three-dimensional wings and configurations beyond the two-dimensional airfoil study.



\bibliography{ifasdREF,Journals,Conferences,General,AePW3}

@book{bisplinghoff-ashley-1962,
	author = {Bisplinghoff, Raymond L. and Ashley, Holt},
	year = {1962},
	publisher = {Wiley \& sons},
	address = {New York},
	title = {Principles of Aeroelasticity}
}

@techreport{theodorsen1949,
	title={General Theory of Aerodynamic Instability and the Mechanism of Flutter},
	author={Theodorsen, Theodore},
	year={1949},
	institution={NACA},
	number={TR-496} 
}

@inproceedings{Wang2019,
	title={{Techniques for Improving Neural Network-Based Aerodynamics Reduced-Order Models}},
	author={Wang, Q. and Medeiros, R. R. and Cesnik, C. E. S. and Fidkowski, K. J. and Brezillon, J. and Bleecke, H. M.},
	booktitle={AIAA SciTech 2019 Forum},
	year={2019}
}

@inproceedings{Wang2020,
	title={{Multivariate Recurrent Neural Network Models for Scalar and Distribution Predictions in Unsteady Aerodynamics}},
	author={Wang, Q. and Cesnik, C. E. S. and Fidkowski, K. J.},
	booktitle={AIAA SciTech 2020 Forum},
	address={Orlando, FL},
	year={2020}
}

@article{FidkowskiRoe2010,
	title={{An Entropy Adjoint Approach to Mesh Refinement}},
	author={Fidkowski, K. J. and Roe, P. L.},
	journal={SIAM Journal on Scientific Computing},
	volume={32},
	number={3},
	pages={1261--1287},
	year={2010},
	doi={10.1137/090759057}
}

@article{Suresh2003,
	title={{Lift Coefficient Prediction at High Angle of Attack Using Recurrent Neural Network}},
	author={Suresh, S. and Omkar, S. N. and Mani, V. and Guru Prakash, T.},
	journal={Aerospace Science and Technology},
	volume={7},
	number={8},
	pages={595--602},
	year={2003}
}

@article{Zhang2012,
	title={{Efficient Method for Limit Cycle Flutter Analysis Based on Nonlinear Aerodynamic Reduced-Order Models}},
	author={Zhang, W. and Wang, B. and Ye, Z. and Quan, J.},
	journal={AIAA Journal},
	volume={50},
	number={5},
	pages={1019--1028},
	year={2012}
}

@inproceedings{Winter2014,
	title={{Reduced-Order Modeling of Unsteady Aerodynamic Loads Using Radial Basis Function Neural Networks}},
	author={Winter, M. and Breitsamter, C.},
	booktitle={Deutscher Luft- und Raumfahrtkongress},
	address={Augsburg, Germany},
	year={2014}
}

@article{Mannarino2014,
	title={{Nonlinear Aeroelastic Reduced Order Modeling by Recurrent Neural Networks}},
	author={Mannarino, A. and Mantegazza, P.},
	journal={Journal of Fluids and Structures},
	volume={48},
	pages={103--121},
	year={2014}
}

@article{Faller1995,
	title={{Real-Time Prediction of Unsteady Aerodynamics: Application for Aircraft Control and Manoeuvrability Enhancement}},
	author={Faller, W. E. and Schreck, S. J.},
	journal={IEEE Transactions on Neural Networks},
	volume={6},
	number={6},
	pages={1461--1468},
	year={1995}
}

@article{Ignatyev2015,
	title={{Neural Network Modeling of Unsteady Aerodynamic Characteristics at High Angles of Attack}},
	author={Ignatyev, D. I. and Khrabrov, A. N.},
	journal={Aerospace Science and Technology},
	volume={41},
	pages={106--115},
	year={2015}
}

@article{KouMultiKernel2017,
	title={{Multi-Kernel Neural Networks for Nonlinear Unsteady Aerodynamic Reduced-Order Modeling}},
	author={Kou, J. and Zhang, W.},
	journal={Aerospace Science and Technology},
	volume={67},
	pages={309--326},
	year={2017}
}

@article{Winter2018,
	title={{Nonlinear Identification via Connected Neural Networks for Unsteady Aerodynamic Analysis}},
	author={Winter, M. and Breitsamter, C.},
	journal={Aerospace Science and Technology},
	volume={77},
	pages={802--818},
	year={2018}
}

@article{Park2013,
	title={{Reduced-Order Model with an Artificial Neural Network for Aerostructural Design Optimization}},
	author={Park, K. H. and Jun, S. O. and Baek, S. M. and Cho, M. H. and Yee, K. J. and Lee, D. H.},
	journal={Journal of Aircraft},
	volume={50},
	number={4},
	pages={1106--1116},
	year={2013}
}

@article{Lindhorst2014,
	title={{Efficient Surrogate Modelling of Nonlinear Aerodynamics in Aerostructural Coupling Schemes}},
	author={Lindhorst, K. and Haupt, M. C. and Horst, P.},
	journal={AIAA Journal},
	volume={52},
	number={9},
	pages={1952--1966},
	year={2014}
}

@article{WinterPOD2016,
	title={{Efficient Unsteady Aerodynamic Loads Prediction Based on Nonlinear System Identification and Proper Orthogonal Decomposition}},
	author={Winter, M. and Breitsamter, C.},
	journal={Journal of Fluids and Structures},
	volume={67},
	pages={1--21},
	year={2016}
}

@article{Hochreiter1997,
	title={{Long Short-Term Memory}},
	author={Hochreiter, S. and Schmidhuber, J.},
	journal={Neural Computation},
	volume={9},
	number={8},
	pages={1735--1780},
	year={1997}
}

@article{Bengio1994,
	title={{Learning Long-Term Dependencies with Gradient Descent Is Difficult}},
	author={Bengio, Y. and Simard, P. and Frasconi, P.},
	journal={IEEE Transactions on Neural Networks},
	volume={5},
	number={2},
	pages={157--166},
	year={1994}
}

@article{Kou2021Review,
	title={{Data-Driven Modeling for Unsteady Aerodynamics and Aeroelasticity}},
	author={Kou, J. and Zhang, W.},
	journal={Progress in Aerospace Sciences},
	volume={125},
	pages={100725},
	year={2021}
}

@article{Kou2019,
	title={{Multi-Fidelity Modeling Framework for Nonlinear Unsteady Aerodynamics of Airfoils}},
	author={Kou, J. and Zhang, W.},
	journal={Applied Mathematical Modelling},
	volume={76},
	pages={832--855},
	year={2019},
	doi={10.1016/j.apm.2019.06.034}
}

@article{Feldwisch2025,
	title={{Increased-Order Model for Unsteady Aerodynamic Nonlinearities Using Neural Networks}},
	author={Feldwisch, J. M.},
	journal={Journal of Aircraft},
	doi={10.2514/1.C038330},
	year={2025}
}

@article{Kou2017,
	title={{Layered Reduced-Order Models for Nonlinear Aerodynamics and Aeroelasticity}},
	author={Kou, J. and Zhang, W.},
	journal={Journal of Fluids and Structures},
	volume={68},
	pages={174--193},
	year={2017}
}

@article{Lee2023,
	title={{Deep Residual Neural Network for Predicting Aerodynamic Coefficient Changes with Ablation}},
	author={Lee, D. H. and Lee, D. and Han, S. and Seo, S. and Lee, B. J. and Ahn, J.},
	journal={Aerospace Science and Technology},
	volume={136},
	pages={108207},
	year={2023}
}

@article{Demo2023,
	title={{A DeepONet Multi-Fidelity Approach for Residual Learning in Reduced Order Modeling}},
	author={Demo, N. and Tezzele, M. and Rozza, G.},
	journal={Advanced Modeling and Simulation in Engineering Sciences},
	volume={10},
	number={1},
	pages={12},
	year={2023}
}

@article{He2020,
	title={{Multi-Fidelity Aerodynamic Data Fusion with a Deep Neural Network Modeling Method}},
	author={He, L. and Qian, W. and Zhao, T. and Wang, Q.},
	journal={Entropy},
	volume={22},
	number={9},
	pages={1022},
	year={2020}
}

@article{Zhang2021,
	title={{Multi-Fidelity Deep Neural Network Surrogate Model for Aerodynamic Shape Optimization}},
	author={Zhang, X. and Xie, F. and Ji, T. and Zhu, Z. and Zheng, Y.},
	journal={Computer Methods in Applied Mechanics and Engineering},
	volume={373},
	pages={113485},
	year={2021}
}

@inproceedings{HeSchumann2020,
	title={{A Framework for the Analysis of Deep Neural Networks in Autonomous Aerospace Applications Using Bayesian Statistics}},
	author={He, Y. and Schumann, J.},
	booktitle={Workshop on Assured Autonomous Systems},
	year={2020}
}

@article{Yan2023,
	title={{Flight Load Calculation Using Neural Network Residual Kriging}},
	author={Yan, Q. and Wan, Z. and Yang, C.},
	journal={Aerospace},
	volume={10},
	number={7},
	pages={599},
	year={2023}
}

@article{Dong2025,
	title={{Control-Oriented Six-Degree-of-Freedom Modeling of Distributed Propulsion Vehicle}},
	author={Dong, Z. and Liu, B. and Da, X. and Yu, W.},
	journal={The Aeronautical Journal},
	volume={129},
	number={1339},
	pages={2460--2479},
	year={2025}
}
\bibliographystyle{ifasd}

\section*{Copyright statement}
The authors confirm that they, and/or their company or organisation, hold copyright on all of the original material included in this paper. The authors also confirm that they have obtained permission from the copyright holder of any third-party material included in this paper to publish it as part of their paper. The authors confirm that they give permission, or have obtained permission from the copyright holder of this paper, for the publication and public distribution of this paper as part of the IFASD 2026 proceedings or as individual off-prints from the proceedings.

\end{document}